\documentclass[fleqn,usenatbib]{mnras}
\usepackage{newtxtext,newtxmath}

\usepackage[T1]{fontenc}
\usepackage{ae,aecompl}

\usepackage{graphicx}	
\usepackage{amsmath}	
\usepackage{amssymb}	
\usepackage{xcolor}
\usepackage{lscape}
\usepackage{numprint}

\newcommand{\NRV}{122}

\newcommand{\Nstar}{TOI-125} 
\newcommand{\Nplanets}{TOI-125b, TOI-125c and TOI-125d} 
\newcommand{\NstarTOI}{TOI-125} 

\newcommand{\vmag}{$11.02 \pm 0.01$} 
\newcommand{\vmagshort}{11.0} 
\newcommand{\VSini}{$1.0\pm0.5$} 
\newcommand{\age}{$6.8 \pm 4.3$} 
\newcommand{\NstarTeff}{$5320\pm39$} 
\newcommand{\FeH}{$-0.02\pm0.03$} 
\newcommand{\NstarRad}{$0.848\pm0.011$} 
\newcommand{\Nstarlogg}{$4.516 \pm 0.024$} 
\newcommand{\Nstardense}{$1.99^{+0.13}_{-0.11}$} 
\newcommand{\NstarMass}{$0.859^{+0.044}_{-0.038}$} 
\newcommand{\spectraltype}{K0V}

\newcommand{\Ndist}{$111.40 \pm 0.44 $} 
\newcommand{\Av}{$0.032 ^{+0.032}_{-0.023}$} 

\newcommand{\LSO}{La Silla Observatory}

\newcommand{\tess}{{\it TESS}}

\newcommand{\gaia}{{\it Gaia}}

\newcommand{\exofast}{{EXOFASTv2}}

\providecommand{\bjdtdb}{\ensuremath{\rm {BJD_{TDB}}}}

\providecommand{\msun}{\ensuremath{\mathrm M_{\sun}}}
\providecommand{\rsun}{\ensuremath{\mathrm R_{\sun}}}
\providecommand{\lsun}{\ensuremath{\mathrm L_{\sun}}}

\providecommand{\me}{\ensuremath{\mathrm M_{\rm E}}}
\providecommand{\re}{\ensuremath{\mathrm R_{\rm E}}}
\providecommand{\fave}{\langle F \rangle}
\providecommand{\fluxcgs}{10$^9$ erg s$^{-1}$ cm$^{-2}$}

\newcommand{\kms}{km\,s$^{-1}$}

\newcommand{\ms}{m\,s$^{-1}$}

\newcommand{\cmss}{cm\,s$^{-2}$}

\newcommand{\gccc}{g\,cm$^{-3}$}
\newcommand{\masy}{mas\,yr$^{-1}$}

\newcommand{\teff}{$T_{\rm eff}$}
\providecommand{\feh}{$\left[{\rm Fe / H}\right]$}

   \title[TOI-125]{Mass determinations of the three mini-Neptunes transiting TOI-125}

\author[L. D. Nielsen et al.]{
L.D. Nielsen,$^{1}$\thanks{E-mail: Louise.Nielsen@unige.ch}
D. Gandolfi,$^{2}$ 
D.J. Armstrong,$^{3,4}$ 
J.S. Jenkins,$^{5}$ 
M. Fridlund,$^{6,7}$ 
\newauthor
N.C. Santos,$^{8,9}$ 
F. Dai,$^{10}$ 
V. Adibekyan,$^{8}$ 
R. Luque,$^{11,12}$ 
J.H. Steffen,$^{13}$ 
M. Esposito,$^{14}$ 
\newauthor
F. Meru,$^{3,4}$ 
S. Sabotta,$^{14}$ 
E. Bolmont,$^{1}$ 
D. Kossakowski,$^{15}$ 
J.F. Otegi,$^{1,16}$ 
F. Murgas,$^{11,12}$ 
\newauthor
M. Stalport,$^{1}$ 
F.~Rodler,$^{17}$ 
M.R. D\'iaz,$^{5}$ 
N.T.~Kurtovic,$^{5}$ 
G. Ricker,$^{18}$ 
R. Vanderspek,$^{18}$ 
\newauthor
D.W. Latham,$^{19}$ 
S. Seager,$^{18,20,21}$ 
J.N.\ Winn,$^{22}$ 
J.M.\ Jenkins,$^{23}$  
R. Allart,$^{1}$ 
\newauthor
J.M.~Almenara,$^{1}$ 
D. Barrado,$^{24}$ 
S.C.C. Barros,$^{8}$ 
D. Bayliss,$^{3,4}$ 
Z.M. Berdi\~{n}as,$^5$ 
\newauthor
I. Boisse,$^{25}$ 
F. Bouchy,$^{1}$ 
P. Boyd,$^{26}$ 
D.J.A. Brown,$^{3,4}$ 
E.M. Bryant,$^{3,4}$ 
C. Burke,$^{18}$ 
\newauthor
W.D. Cochran,$^{27}$ 
B.F. Cooke,$^{3,4}$ 
O.D.S. Demangeon,$^{8}$ 
R.F. D{\'i}az,$^{28, 29}$ 
J. Dittman,$^{20}$ 
\newauthor
C. Dorn,$^{6}$ 
X. Dumusque,$^{1}$ 
R. A. Garc\'ia,$^{30,31}$ 
L. Gonz\'alez-Cuesta,$^{11,12}$ 
S. Grziwa,$^{32}$ 
\newauthor
I. Georgieva,$^{7}$ 
N. Guerrero,$^{18}$ 
A.P. Hatzes,$^{14}$ 
R. Helled,$^{6}$ 
C.E. Henze,$^{23}$ 
\newauthor
S. Hojjatpanah,$^{8,9}$ 
J. Korth,$^{32}$ 
K.W.F.~Lam,$^{33}$ 
J.~Lillo-Box,$^{24}$ 
T.A. Lopez$^{25}$, 
\newauthor
J. Livingston,$^{34}$ 
S. Mathur,$^{11,12}$ 
O. Mousis$^{25}$, 
N. Narita,$^{11,35,36,37}$ 
H.P. Osborn,$^{25,38}$ 
\newauthor
E. Palle,$^{11,12}$ 
P.A. Pe{\~n}a Rojas,$^5$ 
C.M.~Persson,$^{7}$ 
S.N. Quinn,$^{19}$ 
H. Rauer,$^{39,33,40}$ 
\newauthor
S. Redfield,$^{41}$ 
A.~Santerne,$^{25}$ 
L.A. dos Santos,$^{1}$ 
J.V. Seidel,$^{1}$ 
S.G. Sousa,$^{8}$ 
\newauthor
E.B. Ting,$^{23}$ 
M. Turbet,$^{1}$ 
S. Udry,$^{1}$ 
A. Vanderburg,$^{42}$ 
V. Van Eylen,$^{43}$ 
J.I. Vines,$^5$ 
\newauthor
P.J. Wheatley$^{3,4}$ 
and P.A. Wilson$^{3,4}$ 
\\
Author affiliations are listed in Appendix \ref{sec:affil}.
}

\date{Accepted January 2020}

\pubyear{2020}

\begin{document}
\label{firstpage}
\pagerange{\pageref{firstpage}--\pageref{lastpage}}
\maketitle

\begin{abstract}
The Transiting Exoplanet Survey Satellite, \tess, is currently carrying out an all-sky search for small planets transiting bright stars. In the first year of the \tess\ survey, steady progress was made in achieving the mission's primary science goal of establishing bulk densities for 50 planets smaller than Neptune. During that year, \tess's observations were focused on the southern ecliptic hemisphere, resulting in the discovery of three mini-Neptunes orbiting the star TOI-125, a V=\vmagshort\ K0 dwarf. We present intensive HARPS radial velocity observations, yielding precise mass measurements for \Nplanets. TOI-125b has an orbital period of 4.65 days, a radius of $2.726 \pm 0.075$ \re, a mass of $ 9.50 \pm 0.88$~\me\ and is near the 2:1 mean motion resonance with TOI-125c at 9.15 days.  TOI-125c has a similar radius of $2.759 \pm 0.10$~\re\ and a mass of $ 6.63 \pm 0.99$~\me, being the puffiest of the three planets.  TOI-125d, has an orbital period of 19.98 days and a radius of $2.93 \pm 0.17$~\re\ and mass $13.6 \pm 1.2$~\me. For TOI-125b and d we find unusual high eccentricities of $0.19\pm 0.04$ and $0.17^{+0.08}_{-0.06}$, respectively. Our analysis also provides upper mass limits for the two low-SNR planet candidates in the system; for TOI-125.04 ($R_P=1.36$~\re, $P=$0.53 days) we find a $2\sigma$ upper mass limit of 1.6~\me, whereas TOI-125.05 ( $R_P=4.2^{+2.4}_{-1.4}$~\re, $P=$ 13.28 days) is unlikely a viable planet candidate with upper mass limit 2.7~\me. We discuss the internal structure of the three confirmed planets, as well as dynamical stability and system architecture for this intriguing exoplanet system. 

\end{abstract}



\section{Introduction}

The Transiting Exoplanet Survey Satellite \citep[\tess\ -][]{Ricker:2015} is more than halfway through a survey of about 85\% of the sky.  More than \numprint{1000} planet candidates have been announced so far. The Level-1 mission goal of \tess, to measure the masses and radii of at least 50 exoplanets smaller than 4~\re. Among the first planets that meet the Level-1 requirement are
HD 15337b \& c \citep[TOI-402,][]{2019ApJ...876L..24G,2019A&A...627A..43D}, 
HD 21749b \citep[TOI-186, GJ 143,][]{2019ApJ...875L...7D,2019A&A...622L...7T},
GJ 357 b \citep[TOI-562,][]{2019A&A...628A..39L},
LTT 1445Ab \citep{2019AJ....158..152W},
HD 23472 b\&c \citep[TOI-174,][]{2019A&A...622L...7T} and
$\pi$ Men c \citep[HD 39091,][]{2018ApJ...868L..39H,2018A&A...619L..10G}.

\tess\ is building on top of a great legacy from {\it Kepler} \citep{2010Sci...327..977B} which detected numerous multi-planet systems for which system architecture has been studied in detail; eg. \citet{Lissauer:2011}. The identification of the distinct populations of Super-Earths and mini-Neptunes separated by a valley caused by stellar irradiation evaporating the planet atmosphere \citep{Fulton:2018,Fulton:2017,2017ApJ...847...29O} is also owed to {\it Kepler}. This process can potentially strip a the planet down to it's core. Multi-planet systems provide prime target for testing both bulk composition models and atmospheric evaporation, and are thus crucial for advancing exoplanet science. 

We present the confirmation and precise mass measurements of three mini-Neptunes orbiting the bright (V=\vmagshort\,mag) K0 dwarf star \Nstar\, see Table \ref{tab:stellar} for a full summary of the stellar properties. This work builds largely on intensive radial velocity follow-up observations with HARPS \citep{HARPS}. The three planets all fall withing the \tess\ Level-1 mission goal, with similar radii but quite different masses. The system was previously validated by \citet{2019AJ....158..177Q}, so the main focus of this paper is the mass characterisation presented in Section \ref{sec:analysis}, analysis of the system architecture in Section \ref{sec:dyn} and internal structure Section \ref{sec:internal}. Finally we explore future possibilities for atmospheric characterisation in Section \ref{sec:atmos}.

\begin{table}
\caption{\label{tab:stellar} Stellar properties for \Nstar.}
\resizebox{\columnwidth}{!}{%
	\begin{tabular}{lcc}
	\hline\hline
	\noalign{\smallskip}
	Property	&	Value	&	Source\\
	\noalign{\smallskip}
	\hline
    \noalign{\smallskip}
    \multicolumn{3}{l}{Other Names}\\
    2MASS ID	&  J01342273-6640328	& 2MASS \\
    \multicolumn{2}{l}{Gaia ID ~~~~~~~~~~~4698692744355471616}	& Gaia DR2 \\
    TIC  ID & 52368076 & \tess \\
    TOI & \NstarTOI & \tess \\
    \\
    \multicolumn{3}{l}{Astrometric Properties}\\
    R.A.		&	 01:34:22.43 	& \tess	\\
	Dec			&	 -66:40:34.8	& \tess	\\
    $\mu_{{\rm R.A.}}$ (\masy) & -119.800  $\pm$0.066 & Gaia DR2 \\
	$\mu_{{\rm Dec.}}$ (\masy) & -122.953  $\pm$0.080 & Gaia DR2\\
    Parallax  (mas) & 8.9755 $\pm$0.0356 & Gaia DR2\\
    Distance  (pc) & \Ndist & Gaia DR2\\
    \\
    \multicolumn{3}{l}{Photometric Properties}\\
	V (mag)		&\vmag  	&Tycho \\
	B (mag)		&11.72 $\pm$0.12	&Tycho\\
    G (mag)		& 10.718 $\pm$ 0.020	&{Gaia}\\
    T (mag)	    & 10.1985 $\pm$ 0.006	&\tess\\
    J (mag)		& 9.466  $\pm$ 0.021	&2MASS\\
   	H (mag)		&  9.112 $\pm$ 0.025	&2MASS\\
	K$_{\rm s}$ (mag) &8.995$\pm$ 0.021	&2MASS\\
    W1 (mag) & 8.945 $\pm$ 0.030	&WISE\\
    W2 (mag) & 9.006 $\pm$ 0.030	&WISE\\
    W3 (mag) & 8.944 $\pm$ 0.030 & WISE\\
    W4 (mag) & 8.613 $\pm$ 0.262& WISE \\
    A$_{V}$	& \Av & Sec. \ref{sec:exofast}\\
    
    \\
    \multicolumn{2}{l}{Bulk Properties}& This work:\\
    \teff\,(K)    & \NstarTeff    &Sec. \ref{sec:spec} \& \ref{sec:exofast}\\
    Spectral type & \spectraltype &Sec. \ref{sec:spec} \& \ref{sec:exofast}\\
    log g (\cmss)& \Nstarlogg & Sec. \ref{sec:exofast}\\
    $\rho$ (\gccc)& \Nstardense & Sec. \ref{sec:exofast}\\
    $\mathrm{\left[Fe/H\right]}$  & \FeH &Sec. \ref{sec:spec} \& \ref{sec:exofast}\\
	{\it v}\,sin\,{\it i} (\kms)	& < \VSini	& Sec. \ref{sec:spec} \\
	Age	(Gyrs) & \age	&	Sec. \ref{sec:exofast}\\
    Radius ($R_{\odot}$) &	\NstarRad & Sec. \ref{sec:exofast}\\
    Mass ($M_{\odot}$) &  \NstarMass & Sec. \ref{sec:exofast}\\
    \noalign{\smallskip}
	\hline
	\noalign{\smallskip}
    \end{tabular}
    }
 Tycho \citep{Tycho}; 2MASS \citep{2MASS}; WISE \citep{WISE}; Gaia \citep{Gaia2018} 
\end{table}


\section{Observations}
\subsection{TESS photometry}
\Nstar\ (TIC 52368076) was observed by \tess\ in Sectors 1 and 2 from 25 July to 20 September 2018. It appeared on CCD1 of camera 3 in Sector 1 and CCD2 of camera 3 in Sector 2. 

The data are available with 2-min time sampling (cadence) and were processed by the Science Processing Operations Center \citep[SPOC - ][]{Jenkins:2016} to produce calibrated pixels, and light curves. Based on the Data Validation report produced by the transit search conducted by the SPOC \citep{Li:DVmodelFit2019,Twicken:DVdiagnostics2018}, two TESS objects-of-interest, TOI-125b and TOI-125c, were announced by the TESS Science Office (TSO) from Sector 1. This was the first multi-planet-candidate system announced by the TSO. With data from Sector 2 a third planet candidate, TOI-125d, was revealed with one transit observed in each sector.

\begin{figure*}
\begin{minipage}{\textwidth}
   \includegraphics[width=\textwidth]{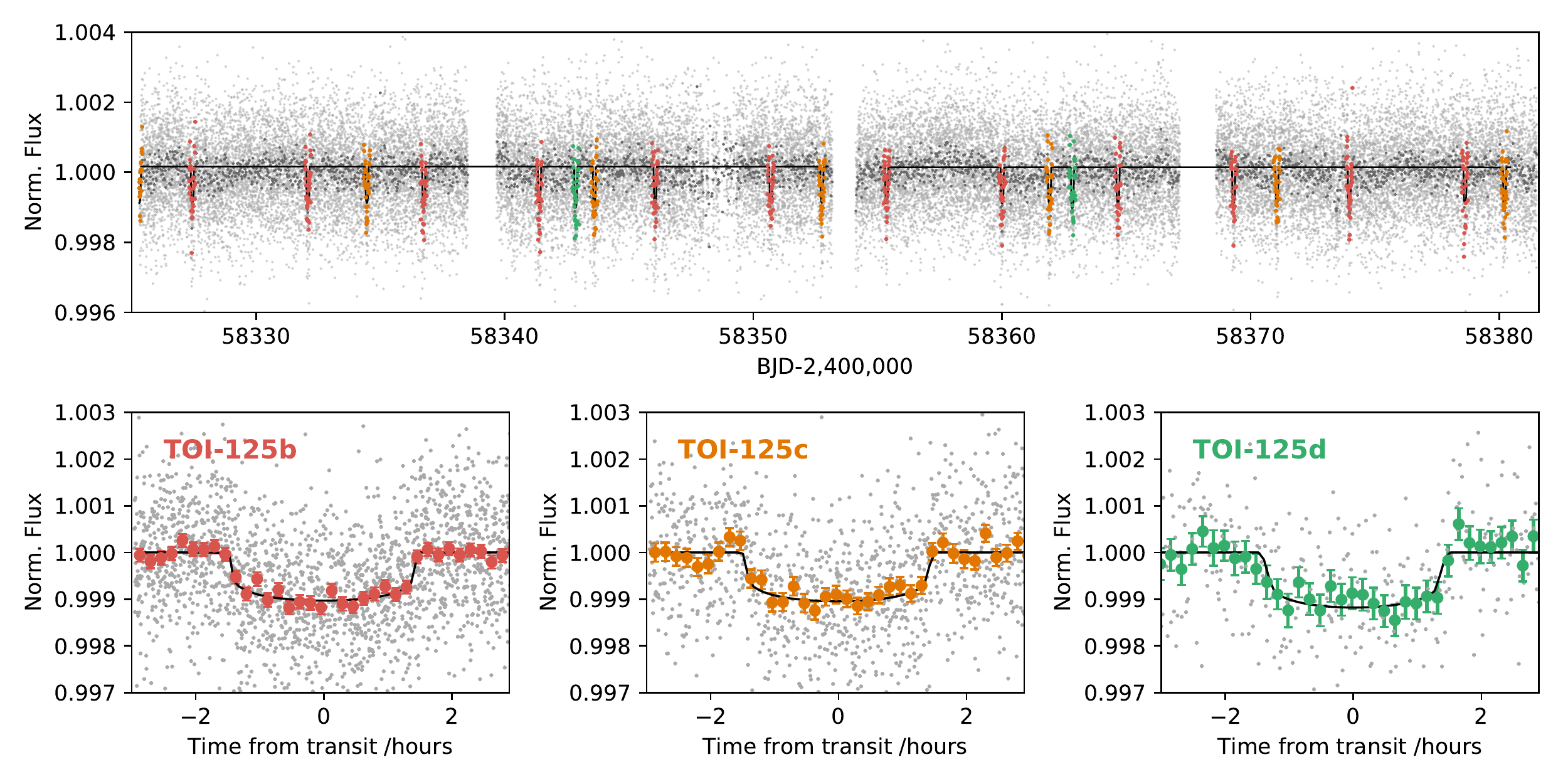}
      \caption{\label{fig:LCs} \tess\ data for TOI-125 spanning Sectors 1 and 2. {\it Top panel:} full light curve with the 2-minute cadence data in light grey and the same data binned to 10 min in dark grey. The binned data surrounding the transits are highlighted in red, yellow and green.
      The light curve from the two Sectors consist of four segments that each correspond to one \tess\ orbit of 13.7 days. After each orbit the spacecraft interrupts observations to downlink the data to Earth, causing gaps in the data coverage. Furthermore, there are features in the light curve from the momentum dumps of the satellite, which take place approximately every 2.5 days. None of the detected transits occurred during momentum dumps.  {\it Bottom panel:} phase folded \tess\ light curves for \Nstar b, \Nstar c and \Nstar d, again with 2-minute cadence data in grey and binned to 10 minutes in the same colours as the top panel.}
\end{minipage}
\end{figure*}

For transit modelling, we used the publicly available Simple Aperture Photometry flux, after the removal of artefacts and common trends with the Pre-search Data Conditioning (PDC-SAP) algorithm \citep{Twicken:2010,2012PASP..124.1000S,2014PASP..126..100S} provided by SPOC. The light curve precision in both sectors is 125\,ppm, averaged over half an hour, consistent with the value predicted by \citet{Sullivan:2015} for a star with apparent \tess-magnitude 10.2. Figure \ref{fig:LCs} shows the full 2 min cadence \tess\ light curve, with data points binned to 10 min over-plotted, along with the phase folded light curves for \Nplanets.

The \Nstar\ system was vetted by \cite{2019AJ....158..177Q} using ground-based photometry, high-angular-resolution imaging and reconnaissance spectroscopy. TOI-125b and TOI-125c were statistically validated as planets, while TOI-125d (then called TOI-125.03) remained a high-SNR planet candidate, based on only two observed transits. Two additional low-SNR candidates were identified: TOI-125.04, with period 0.53 days making it an ultra short period (USP) planet candidate and TOI-125.05 at 13.28 days. \cite{2019AJ....158..177Q} stressed that these two candidates are marginal detections, and did not attempt to validate them statistically.

\subsection{High resolution spectroscopy with HARPS}
\Nstar\ was observed intensively with the HARPS spectrograph \citep{HARPS} on the ESO 3.6m telescope at \LSO, Chile, from 21 September 2018 to 8 January 2019. In total 122 spectra were obtained under programmes 1102.C-0249 (PI: Armstrong), 0101.C-0829/1102.C-0923 (PI: Gandolfi), 0102.C-0525 (PI: D\'iaz), 0102.C-0451 (PI: Espinoza) and 60.A-9700 (technical time).
HARPS is a stabilised high-resolution spectrograph with resolving power $R\sim\numprint{115000}$, capable of sub-\ms radial velocity (RV) precision. We used the instrument in high-accuracy mode with a $1\arcsec$ science fibre on the star and a second fibre on sky to monitor the sky-background during exposure. We used a nominal exposure time of 1800 seconds, which on occasion was adjusted within a range of 800 -- 2100 seconds depending on sky-condition and observation schedule.

RVs were determined with the standard (offline) HARPS data reduction pipeline using a K0 binary mask for the cross correlation \citep{Pepe:2002}, and a K3 template for flux correction to match the slope of the spectra across echelle orders. We performed the data reduction uniformly for all the data from the 6 programmes under which data had been acquired, to mitigate any possible RV offsets induced by different data reduction parameters and catalogue-coordinates in the FITS headers. With a typical signal-to-noise ratio (SNR) of 55, we achieved an RV precision of 1.5 \ms. The RV data have been made publicly available through The Data \& Analysis Center for Exoplanets (DACE\footnote{\url{https://dace.unige.ch/radialVelocities/?pattern=TOI-125}}) hosted at the university of Geneva.
For each epoch the bisector-span (BIS), contrast and FWHM of the CCF were calculated, as well as the chromospheric activity indicators Ca II H\&K, $\mathrm{H\alpha}$, and Na. 

In our RV analysis we excluded data taken on the nights starting 25, 26 and 27 November 2018. On these dates the ThAr lamp used for wavelength calibration of HARPS was deteriorating and subsequently exchanged on 28 November 2018\footnote{See HARPS instrument monitoring pages: \url{https://www.eso.org/sci/facilities/lasilla/instruments/harps/inst/monitoring/thar_history.html}}. The changing flux ratio between thorium and argon emission lines of the dying ThAr lamp induced a $2\,\mathrm{m\,s^{-1}\,day^{-1}}$ drift in the wavelength solution of HARPS over 5 days. 
The problematic data were confirmed by comparing unpublished data from the \mbox{HARPS-N} solar telescope \citep{Dumusque:2015,CollierCameron:2019} and Helios on HARPS, which also observes the Sun daily. The Helios RVs show a clear drift away from the RVs from the HARPS-N solar telescope 25 - 27 November 2018, before returning to a nominal level after the change of the ThAr lamp. We still include spectra taken on those dates in our spectral analysis described in the following Section ~\ref{sec:spec}.

We clearly detect RV signals for \Nplanets. The top panel in Figure~\ref{fig:periodogram} shows a Lomb-Scargle periodogram of the raw RVs in which there are clear signals at 4.65 and 19.98 days for TOI-125b and TOI-125d with False-alarm-probability (FAP) < 0.1\%. A hint can be seen for TOI-125c at 9.15 days, possibly interfering with a $P/2$ alias from TOI-125d. The residuals of a two-planet fit is shown in Figure~\ref{fig:periodogram} below the raw RVs, and show a significant peak at the period of TOI-125c. Peaks at 1.27 and 0.82 days in the raw RVs are aliases of TOI-125b. No signal is found for TOI-125.04 ($P=0.53$~days) nor TOI-125.05 ($P=13.28$~days).

\section{Data analysis and results} \label{sec:analysis}
\subsection{Spectral classification and stellar chemical abundances} \label{sec:spec}
The \NRV\ 1-D HARPS spectra were stacked to produce a high fidelity spectrum with SNR per resolution element $\sim$500 at 5500\,\AA\ for spectral analysis.  Retrieving stellar parameters from the observed spectrum can be done using several different methods.  In the case of \Nstar, stellar atmospheric parameters (\teff, \feh, $\log g$) and relative abundances of refractory material were derived using two different methodologies: a) as described in \cite{Sousa:2008} and \cite{Santos:2013} 
using equivalent widths (EW) of chosen lines while assuming ionisation- and excitation equilibrium, and b) with the Spectroscopy Made Easy (SME) code \citep{1996A&AS..118..595V, 2005ApJS..159..141V, 2017A&A...597A..16P} as applied to a grid of model atmospheres. 

For the first method, \teff, \feh, and $\log g$ were calculated using the EW of 237 FeI and 33 FeII lines. A grid of Kurucz model atmospheres \citep{Kurucz:1993} and the radiative transfer code MOOG \citep{Sneden:1973} were used to model the stellar atmosphere. For the derivation of abundances of refractory elements we used the approach from \cite{2015A&A...583A..94A}. \Nstar\ shows typical abundances for a main sequence star, comparable to the ensemble of HARPS GTO stars.

As a second approach we used SME version 5.22 applied to a grid of MARCS model atmospheres. These are 1D-LTE plane-parallel and spherically-symmetric model atmospheres applicable to solar-like stars \citep{2008A&A...486..951G}. Synthetic spectra were then calculated based on the model grid and fit to the observed spectral features, 
focusing on those that are especially sensitive to different photospheric parameters, including \teff, \feh, $\log g$, micro- and macro turbulence and rotational velocity ($v \sin i$). Here one is changing one or more input parameters and then iteratively using a $\chi ^2$ minimisation procedure to arrive at the actual stellar parameters. We used the calibration equation of \cite{2010MNRAS.405.1907B} and \cite{2014MNRAS.444.3592D} to estimate the micro- and macro- turbulent velocities, based on the derived values on \teff\ and $\log g$. We also fitted 45 isolated and unblended metal lines to determine the projected stellar rotation velocity ($v \sin i$), 
which was found to be 1.0$\pm$0.5 km/s. 

The derived parameters and abundances for both methods are presented in Table \ref{tab:spec}. It should be noted that the uncertainties were derived from internal errors only, and thus do not include uncertainties inherent to the models themselves. 
While the abundances and surface gravity $\log g$ agree as a whole, there is a $2\sigma$ discrepancy between the effective temperature, \teff, obtained with the two methods. The \feh\ measurements also differ slightly between the two methods, but are consistent to $1\sigma$. We have investigated the impact of this on the final set of system parameters, and found less than $5\%$ difference in stellar and planetary masses and radii. For the final modelling of the system, we used the average of \teff\ and \feh\ as Gaussian priors in the MCMC. The errors were inflated to encompass both values at a $1\sigma$ level, in order to reflect the model dependency of the stallar atmospheric parameters. 

\begin{table}
\centering
  \caption{ \label{tab:spec}Spectral parameters derived from the stacked HARPS spectrum with SNR/resolution element $\sim$500 at 5500\AA, using two different methods. \teff\ and \feh\ and their uncertainties were used as Gaussian priors on the MCMC joint modelling of the planetary and stellar parameters - we used the average between the two approaches. The errors were inflated to encompass both values at a $1\sigma$ level, in order to reflect the model dependency of the atmospheric parameters. $V_t$ denotes micro- and macro turbulence velocities.}
 \begin{tabular}{lcccc}
  \hline
  \hline
  \noalign{\smallskip}
  &\multicolumn{2}{c}{Equivalent width} &\multicolumn{2}{c}{SME} \\
  Parameter & Value & 1$\sigma$  & Value & 1$\sigma$ \\
  \noalign{\smallskip}
  \hline
  \noalign{\smallskip}
  \teff (K)&    5295 &      42 & 5125 & 60\\
  $\log g$ (cgs) & 4.51 & 0.07 & 4.4 & 0.2\\
  $V_t \mathrm{micro}$ (\kms)     &0.72  & 0.09& 0.8 & 0.1 \\
  $V_t \mathrm{macro}$ (\kms)     &     &    & 2.5 & 0.5 \\
  {\it v}\,sin\,{\it i} (\kms) & & & 1.0 & 0.5 \\
  \feh (dex) & -0.02 & 0.03  & 0.00 & 0.05 \\
  NaI/H	(dex)&		$-$0.06	& 0.05	& $-$0.1	& 0.05 \\	
MgI/H	(dex)&		0.01	& 0.05	& 	&  \\
AlI/H	(dex)&		$-$0.02	& 0.07	& 	&  \\
SiI/H	(dex)&		$-$0.04	& 0.06	& $-$0.1	& 0.05 \\
CaI/H	(dex)&		$-$0.03	& 0.07	& $-$0.1	& 0.05 \\
ScII/H  (dex)&		$-$0.02	& 0.04	& 	    &      \\
TiI/H	(dex)&		0.09	& 0.06	& $-$0.05	& 0.05 \\
CrI/H	(dex)&		0.02	& 0.05	& 0.0	& 0.05 \\
NiI/H	(dex)&		$-$0.06	& 0.03	& $-$0.05	& 0.05 \\
Zr/H	(dex)&				&       & $-$0.1	& 0.05 \\	
\noalign{\smallskip}
\hline
 \end{tabular}
\end{table}

\subsection{Stellar rotation and activity}
The average value of the Ca {\sc ii} H\,\&\,K chromospheric activity indicator for \Nstar\ is $\log R^{\prime}_\mathrm{HK}\,=\,-5.00\,\pm\,0.08$, indicating a low activity level that would introduce an RV-signal on the scale of 0.4 \ms\ \citep{Suarez:2017}. According to \cite{Suarez:2015}, the expected rotation period of an early K-type dwarf with $\log R^{\prime}_\mathrm{HK}\,=\,-5.00\,\pm\,0.08$ is $P_\mathrm{rot}\,=\,32^{+5}_{-4} ~\mathrm{days}$. This is in good agreement with the classical empirical relation from \cite{Noyes:1984}, which gives $P_\mathrm{rot}\,=\,31\,\pm\,6~\mathrm{days}$. Assuming that the star is seen equator-on, the projected rotational velocity $v \sin i = 1 \pm 0.5$ \kms\ and stellar radius imply a rotation period of $ \lesssim 43$~days. This could be indicative of the stellar spin and the planetary orbits being aligned.

\begin{figure*}
\begin{minipage}{\textwidth}
\includegraphics[width=\columnwidth,trim={0cm 2.7cm 14.0cm 6.5cm},clip]{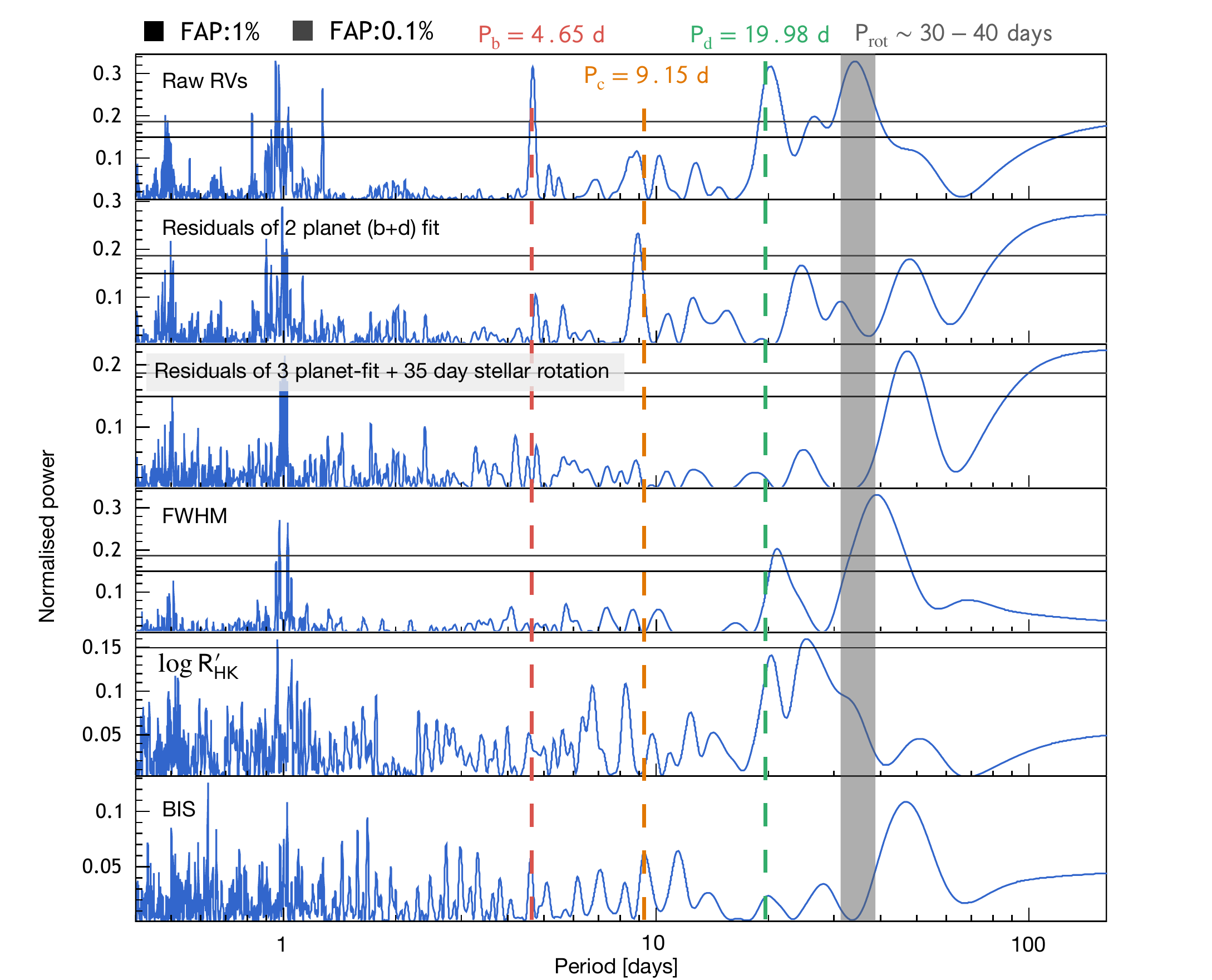}
\caption{\label{fig:periodogram} Lomb-Scargle periodograms, from the top: raw RVs, residuals of a 2 planet fit (including TOI-125b and d), FWHM, $\log R^{\prime}_{HK} $ and bisector span. 1\% and 0.1\%  FAP are indicated as horizontal lines. Orbital periods for \Nplanets\ are marked as red, yellow and green dashed lines. The expected rotational period of the star is highlighted in grey.}
\end{minipage}
\end{figure*}

We searched the RVs and activity indicators for a signal matching the expected $P_\mathrm{rot}$. Figure~\ref{fig:periodogram} shows Lomb-Scargle periodograms derived for the raw RVs, RV-residuals to a two-planet fit and RV-residuals to a three-planet fit including an additional term fitting a possible 35 days period. We also include periodograms of, FWHM, $\log R^{\prime}_{HK} $ and bisector span. False-alarm-probability (FAP) thresholds have been computed analytically for levels of 1\% and 0.1\%. For both the RVs and FWHM there is a FAP > 0.1\% signal close to 40 days, highlighted in grey in Figure \ref{fig:periodogram}. This is in reasonable agreement with the expected stellar rotation period based on $\log R^{\prime}_{HK}$. The periodogram for $\log R^{\prime}_{HK}$ has signal at 25.5 days, which could be a $P_{\mathrm{rot}}/2$ alias with FAP 1\%. BIS show no significant signals, though the main peak at 47 days somewhat matches the ones found in FWHM and $\log R^{\prime}_{HK}$. As a test, we fit three planets along with a 35 day modulation mimicking a signal induced by stellar rotation. The periodogram of the residuals is presented in Figure~\ref{fig:periodogram}. It is evident that residual signals at longer periods are still present, including a long term ($P> 100$~days) signal we later model as quadratic drift in the RVs.

We searched both the SAP and PDC-SAP light curves for photometric modulation from stellar rotation, but found no convincing signal. This is not too surprising as the baseline of the \tess\ observation is short (2 x 27 days) compared to the expected rotational period ($\geq 30$~days).

Based on the signal seen in both FWHM and RV measurements at 40 days, we attempted to model our RVs with a Gaussian process (GP) trained on the FWHM using a quasi-periodic kernel. The GP had problems converging and the planets' parameters were unchanged from a classic RV fit. Given the low SNR of both the FWHM and $\log R^{\prime}_{HK}$ indicators, combined with the small expected effect of stellar activity on the RVs, we proceeded to model our data without a GP. Since the period of TOI-125d is about half the stellar rotation period, this might affect the mass measurement of that planet, but we expect this to be a small offset. We can however not exclude that the RV semi-amplitude of TOI-125d is slightly affected by stellar activity. Our RV data spans several stellar rotations, which to some degree helps mitigate this as we average over epochs with different activity levels. 

\subsection{Joint modelling with \exofast} \label{sec:exofast}

The planetary and stellar parameters were modelled self-consistently through a joint fit of the HARPS RVs and \tess\ photometry with \exofast\ \citep{Exofastv2, Exofast}. \exofast\ can fit any number of transits and RV sources for a given number of planets while exploring the vast parameter space through a differential evolution Markov Chain coupled with a Metropolis-Hastings Monte Carlo sampler.

The local $\chi^2$-minimum in parameter space is identified with \texttt{AMOEBA}, which is a non-linear minimiser using a downhill simplex method \citep{NelderMead:1965}. The starting point of the MCMC is set to be within $1\sigma$ of the best-fit value. Hereafter the full parameter space is explored with a Monte Carlo sampler in numerous steps. At each step the stellar properties are modelled, and limb darkening coefficients for this specific star are calculated by interpolating tables from \cite{ClaretBloemen:2011}. The analytic expressions from \cite{MandelAgol:2002} are used for the transit model. The eccentricity is parameterised as $e^{\frac{1}{4}} \cos( \omega_*)$ and $e^{\frac{1}{4}} \sin (\omega_*)$ to impose uniform eccentricity priors and mitigate Lucy-Sweeney bias of final measurement \citep{LucySweeney:1971}. \exofast\ rejects any solutions where the planetary orbits cross.

At each step $\chi^2$ is evaluated and assumed to be proportional to the likelihood, which is true for fixed uncertainties. The Metropolis-Hastings algorithm is invoked and 20\% of all steps with lower likelihood are kept in the chain. The MCMC thus samples the full posterior distribution.

The size and direction of the next step in the MCMC is determined by the differential evolution Markov Chain method \citep{terBraak:2006}, where several chains (twice the number of fitted parameters) are run in parallel. The step is determined by the difference between two random chains. 
In \exofast\, a self adjusting step size scale is implemented to ensure optimal sampling across the orders of magnitude difference in scales of uncertainty. This is crucial to effectively sample all parameters (e.g,  from the orbital period which can be determined to $10^{-4}$ days for transiting planets to the RV semi-amplitude which commonly can have 10\% uncertainty). 

The first part of the chains with $\chi^2$ above the median $\chi^2$ are discarded as the 'burn-in' phase, so as not to bias the final posterior distributions toward the starting point. A built-in Gelman-Rubin statistic \citep{Gelman:1992, Gelman:2003, Ford:2006} is used to check the convergence of the chains. When modelling RVs and transit photometry simultaneously, each planet has seven free parameters and up to four additional RV terms for the systemic velocity, drift of the system, and jitter. For the transit light curve two limb darkening coefficients for the \tess\ band are fitted, along with the baseline flux and variance of the light curve. 

Another four parameters are fitted for the star: \teff, \feh, $\log M_*$, and $R_*$. We applied Gaussian priors on \teff\ and \feh\ from the spectral analysis, presented in Section \ref{sec:spec}. The mean stellar density is determined from the transit light curve. The \gaia\ DR2 parallax was used, along with SED-fitting to constrain the stellar radius further. We include the broad band photometry presented in Table \ref{tab:stellar} in our analysis, apart from the very wide Gaia G-band. We set an upper limit on the V-band extinction from \cite{Schlegel:1998} and \cite{Schlafly} to account for reddening along the line of sight. Combining spectroscopic \teff\ and \feh\ with broad band SED-fitting allow us to perform detailed modelling of the star with the MESA Isochrones and Stellar Tracks \citep[MIST][]{Mist0,Mist1}, which are evaluated at each step in the MCMC.

\begin{figure*}
\begin{minipage}{\textwidth}
  \includegraphics[width=\textwidth,trim={1.5cm 0.4cm 2.5cm 1.5cm},clip]{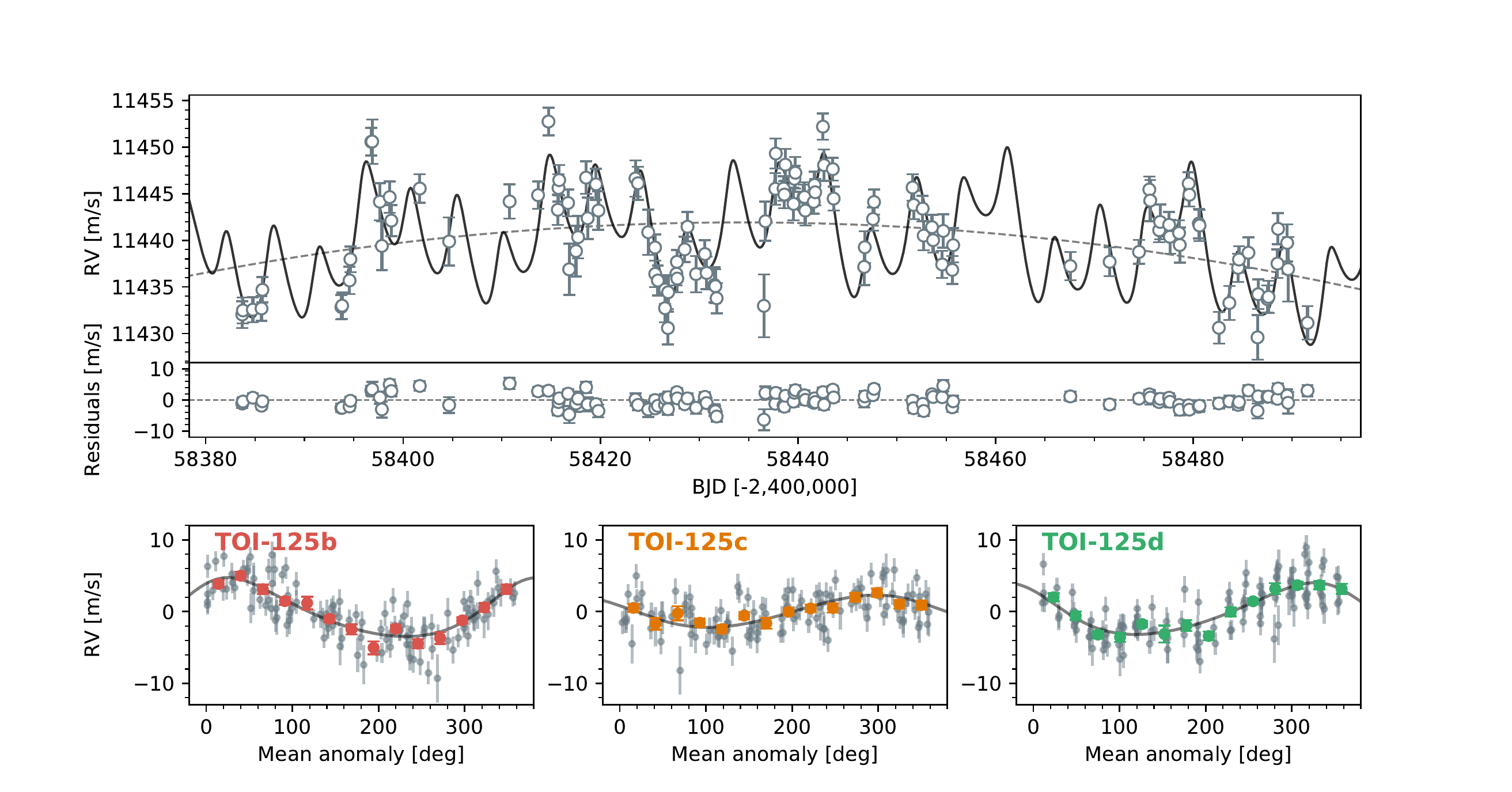}
      \caption{\label{fig:RVs} HARPS RVs for \Nstar\ with a three planet model including eccentric orbits and a quadratic drift. The residuals to the best fit are shown right below the RV timeseries. The bottom panel shows the data phase folded and binned for each planet.}
\end{minipage}
\end{figure*}

We ran \exofast\ with \numprint{50000} steps on the HARPS RVs and \tess\ photometry with a quadratic drift in the RVs, with and without eccentricities for \Nplanets.
TOI-125b and TOI-125d have significant eccentricity. Figure \ref{fig:RVs} displays the HARPS RVs with the final model and Figure~\ref{Hist_eb} shows a sample of the posterior distribution for the eccentricity of TOI-125b. For simplicity we fit eccentricities for all three planets in the system. 

The final median values of the posterior distributions and their $1\sigma$ confidence intervals for the stellar and planetary parameters are listed in Table \ref{tab:TOI-125}. 
We find that TOI-125b has an orbital period of 4.65 days, a radius of $2.726 \pm 0.075$~\re, and a mass of $ 9.50 \pm 0.88$~\me, yielding a mean density of 2.57~\gccc. It has the highest orbital eccentricity of the three planet in the system, $e_b = 0.194 ^{+0.041}_{-0.036}$.
With an orbital period of 9.15~days, TOI-125c is near the 2:1 mean motion resonance with its inner companion. It has a radius of $2.759 \pm 0.10$~\re\ and a mass of $ 6.63 \pm 0.99$~\me, implying a mean density of 1.73~\gccc. TOI-125d is thus the least dense of the three. It's orbital eccentricity is consistent with zero, $e_c = 0.066^{+0.070}_{-0.047}$.
The outer transiting planet, TOI-125d, has an orbital period of 19.98 days and eccentricity $e_d = 0.168^{+0.088}_{-0.062}$. With a radius of $2.93 \pm 0.17$~\re\ and mass $13.6 \pm 1.2$~\me, it is the densest of the three planets, with $\rho_P = 2.98$~\gccc.

\Nplanets\ are thus all mini-Neptunes with similar radii, but different masses yielding a high-low-higher density pattern outwards in the system. The planets straddle the gap identified in the mass-period plane by \cite{Armstrong_2019}. All three planets have the same orbital inclination to within a degree. The high orbital eccentricities detected for TOI-125b and d are unusual for such a compact system of mini-Neptunes \citep{2019AJ....157...61V}.

\Nstar\ is found to be a main-sequence K0-star with mass \NstarMass~ \msun, radius \NstarRad~ \rsun\ and \teff=\NstarTeff~K. This is in reasonable agreement with the properties reported in the Gaia Data Release 2: $ R_* = 0.90 \pm 0.03$ \rsun\ and \teff $= 5150 \pm 84 ~{\rm K}$ \citep{Gaia2018}. The quadratic drift found in the RVs might indicate the existence of an additional massive companion in the system, at a long period $P \gtrsim 100~\mathrm{days}$. We obtained a few RV points in July 2019 with low precision to rule out a stellar companion. More high-precision RVs would be needed to determine the nature of this long-term signal.

\begin{table*}
\begin{minipage}{\textwidth}
\caption{\label{tab:TOI-125} Median values and 68\% confidence intervals for \Nplanets\ and their host star fitted with \exofast, while including a quadratic RV drift and orbital eccentricities for all three planets.}
\begin{tabular}{lcccc}\hline
\smallskip\\\multicolumn{2}{l}{Stellar Parameters:}&\smallskip\\
~~~~$M_*$\dotfill &Mass (\msun)\dotfill &$0.859^{+0.044}_{-0.038}$\\
~~~~$R_*$\dotfill &Radius (\rsun)\dotfill &$0.848\pm0.011$\\
~~~~$L_*$\dotfill &Luminosity (\lsun)\dotfill &$0.519\pm0.016$\\
~~~~$\rho_*$\dotfill &Density (cgs)\dotfill &$1.99^{+0.13}_{-0.11}$\\
~~~~$\log{g}$\dotfill &Surface gravity (cgs)\dotfill &$4.516 \pm 0.024$\\
~~~~$T_{\rm eff}$\dotfill &Effective Temperature (K)\dotfill &$5320 \pm 39 $\\
~~~~$[{\rm Fe/H}]$\dotfill &Metallicity (dex)\dotfill &$-0.02 \pm 0.03$\\
~~~~$Age$\dotfill &Age (Gyr)\dotfill &$6.8^{+4.4}_{-4.1}$\\
~~~~$A_V$\dotfill &V-band extinction (mag)\dotfill &$0.032^{+0.032}_{-0.023}$\\
~~~~$d$\dotfill &Distance (pc)\dotfill &$111.40\pm0.44$\\
~~~~$\dot{\gamma}$\dotfill &RV slope (m/s/day)\dotfill &$-0.0123\pm0.0078$\\
~~~~$\ddot{\gamma}$\dotfill &RV quadratic term (m/s/$\mathrm{day^2}$)\dotfill &$-0.00183\pm0.00025$\\

\smallskip\\\multicolumn{2}{l}{Planetary Parameters:}&b&c&d\smallskip\\
~~~~$P$\dotfill &Period (days)\dotfill
&$4.65382^{+0.00033}_{-0.00031}$&$9.15059^{+0.00070}_{-0.00082}$&$19.9800^{+0.0050}_{-0.0056}$\\
~~~~$R_P$\dotfill &Radius (\re)\dotfill & $2.726 \pm 0.075$ &$2.759 \pm 0.10 $ & $2.93\pm0.17$\\
~~~~$M_P$\dotfill &Mass (\me)\dotfill & $9.50 \pm 0.88 $ & $6.63 \pm 0.99 $ & $13.6 \pm 1.2$\\
~~~~$\rho_P$\dotfill &Density (cgs)\dotfill &$2.57\pm0.33$ & $1.73 \pm 0.33 $ & $2.98^{+0.65}_{-0.52}$\\
~~~~$T_C$\dotfill &Time of conjunction (\bjdtdb)\dotfill & $58355.35529 \pm 0.0010$ & $58361.9085 \pm 0.0013$ & $58342.8516 \pm 0.0039$\\
~~~~$a$\dotfill &Semi-major axis (AU)\dotfill &$0.05186^{+0.00086}_{-0.00077}$ & $0.0814 \pm 0.0013 $& $0.1370 \pm 0.0022$\\
~~~~$b$\dotfill &Transit impact parameter \dotfill & $0.27^{+0.17}_{-0.18}$& $0.522^{+0.086}_{-0.18}$ & $0.652^{+0.093}_{-0.16}$\\
~~~~$i$\dotfill &Inclination (Degrees)\dotfill &$88.92^{+0.71}_{-0.60}$ & $88.54^{+0.41}_{-0.19}$ & $88.795^{+0.18}_{-0.10}$\\
~~~~$e$\dotfill &Eccentricity$^{\dagger}$ \dotfill &$0.194^{+0.041}_{-0.036}$ & $0.066^{+0.070}_{-0.047}$ & $0.168^{+0.088}_{-0.062}$\\
~~~~$\omega_*$\dotfill &Argument of Periastron (Degrees)\dotfill &$-37^{+12}_{-14}$&$70^{+100}_{-110}$& $46^{+23}_{-44}$\\
~~~~$T_{eq}$\dotfill &Equilibrium temperature (K)\dotfill &$1037\pm11$ & $827.8 \pm 8.6$ & $638.1 \pm 6.6$\\
~~~~$\fave$\dotfill &Incident flux (\fluxcgs)\dotfill &$0.252 \pm 0.012 $ & $0.1056 \pm 0.0045$ & $0.0363 \pm 0.0019$\\
~~~~$K$\dotfill &RV semi-amplitude (m/s)\dotfill & $4.11\pm0.36$ & $2.25\pm0.33$ &$3.61\pm 0.31 $ \\
~~~~$R_P/R_*$\dotfill &Radius of planet in stellar radii \dotfill & $0.02950 \pm 0.00070$ &$0.02985 \pm 0.00099 $ & $0.0317\pm0.0018$\\
~~~~$a/R_*$\dotfill &Semi-major axis in stellar radii \dotfill & $13.16 \pm 0.27$ & $20.66 \pm 0.42$ & $34.77 0.70$\\
~~~~$\delta$\dotfill &Transit depth (fraction)\dotfill &$0.000870^{+0.000043}_{-0.000040}$&$0.000891^{+0.000060}_{-0.000057}$&$0.00100^{+0.00012}_{-0.00011}$\\
~~~~$\tau$\dotfill &Ingress/egress duration (days)\dotfill &$0.00380^{+0.00061}_{-0.00026}$&$0.00486^{+0.00079}_{-0.00093}$ & $0.0068^{+0.0021}_{-0.0017}$\\
~~~~$T_{14}$\dotfill &Total transit duration (days)\dotfill &$0.1234\pm0.0024$ & $0.1231^{+0.0026}_{-0.0030}$ & $0.1297^{+0.0070}_{-0.0057}$\\
~~~~$T_{FWHM}$\dotfill &FWHM transit duration (days)\dotfill & $0.1194^{+0.0023}_{-0.0024}$ & $0.1182^{+0.0027}_{-0.0031}$ & $0.1227^{+0.0076}_{-0.0062}$\\
~~~~$T_P$\dotfill &Time of Periastron (\bjdtdb)\dotfill & $58326.03^{+0.17}_{-0.20}$ & $58334.1^{+2.3}_{-2.8}$ & $58341.1^{+1.1}_{-2.4}$\\
~~~~$T_S$\dotfill &Time of eclipse (\bjdtdb)\dotfill & $58325.546^{+0.084}_{-0.081}$ & $58339.06^{+0.30}_{-0.24}$ & $58334.18^{+0.54}_{-0.57}$\\

~~~~$logg_P$\dotfill &Surface gravity \dotfill & $3.097 \pm 0.047$ & $2.931^{+0.068}_{-0.076}$ & $3.192\pm 0.064$\\
~~~~$\Theta$\dotfill &Safronov Number \dotfill & $0.0148^{+0.0014}_{-0.0013}$ & $0.0160\pm0.0024$& $0.0522^{+0.0054}_{-0.0051}$\\

\smallskip\\\multicolumn{2}{l}{Wavelength Parameters:}&TESS\smallskip\\
~~~~$u_{1}$\dotfill &linear limb-darkening coeff \dotfill &$0.382\pm0.035$\\
~~~~$u_{2}$\dotfill &quadratic limb-darkening coeff \dotfill &$0.240^{+0.035}_{-0.036}$\\
\smallskip\\\multicolumn{2}{l}{Telescope Parameters:}& HARPS \smallskip\\
~~~~$\gamma_{\rm rel}$\dotfill &Relative RV Offset (m/s)\dotfill &$11441.90\pm0.30$\\
~~~~$\sigma_J$\dotfill &RV Jitter (m/s)\dotfill &$1.63^{+0.24}_{-0.22}$\\
\smallskip\\\multicolumn{2}{l}{Transit Parameters:}&TESS Sector 1 & TESS Sector 2\smallskip\\
~~~~$\sigma^{2}$\dotfill &Added Variance \dotfill &$-0.000000023\pm0.000000027$&$-0.000000046\pm 0.000000027$\\
~~~~$F_0$\dotfill &Baseline flux \dotfill &$1.000136\pm0.000019$&$1.000151\pm0.000019$\\
\hline
\end{tabular}
~$^{\dagger}$ The eccentricities presented here are the direct outputs from \exofast, without any constraints from N-body simulations. Our dynamical analysis in Sec.~\ref{sec:nbody} puts upper limits on the eccentricities for TOI-125b and TOI-125c, but retains the same eccentricities within a $1-\sigma$ confidence interval.
\end{minipage}
\end{table*}

\subsubsection{Marginal planet candidates TOI-125.04 and TOI-125.05}
\begin{figure}
\includegraphics[width=\columnwidth,trim={0.0cm 0.65cm 0.0cm 0.255cm},clip]{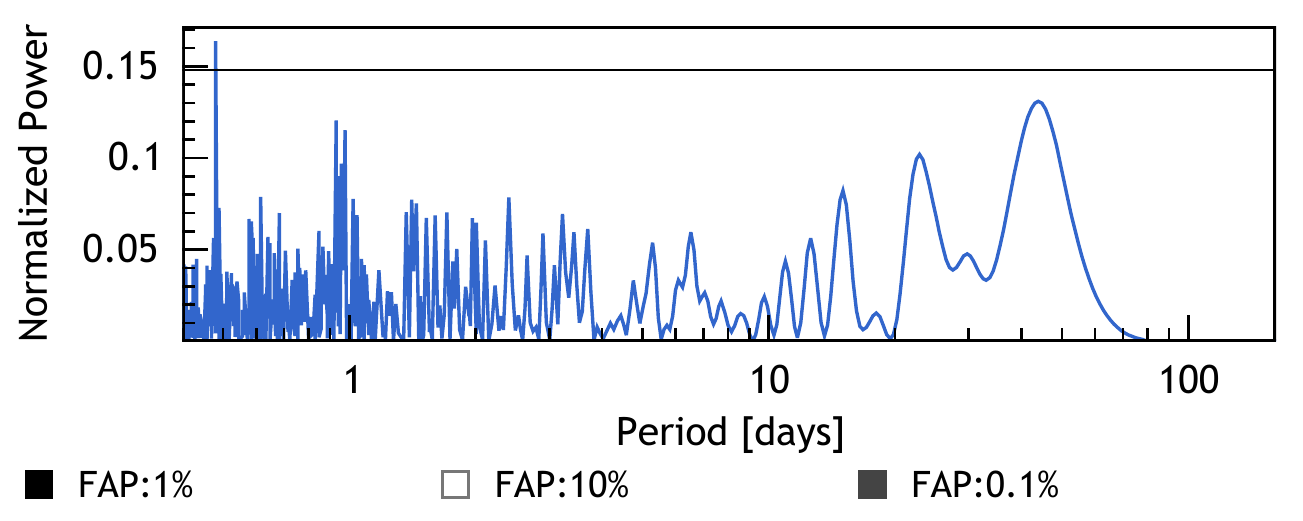}
\caption{\label{fig:LSresiduals} Periodogram of the RV residuals after fitting 3 planets with eccentric orbits and a quadratic trend. The horizontal black line is the 1\% FAP.}
\end{figure}

Figure \ref{fig:LSresiduals} shows the residuals from the 3-planet fit. We see no hint of any signal from TOI-125.04 or TOI-125.05. The strong peak at $P=0.49$ days is an alias of the residual signal at 50 days.

We derive upper mass limits for the two planet candidates by running \exofast\ on the HARPS RVs while only including priors on the orbital period and transit depth from \cite{2019AJ....158..177Q}. We do not include the \tess\ photometry, to save computational time. Fitting 3, 4 or 5 planets has little impact on the final parameters for \Nplanets.
For the marginal USP candidate TOI-125.04 ($P=0.53$ days, $R_P = 1.36^{+0.14}_{-0.16}$ \re) we find a radial velocity semi-amplitude of $K=0.56^{+0.4}_{-0.3}$ \ms corresponding to a $2\sigma$ upper mass limit of 1.6 \me. Our measurement are compatible with no planet and we cannot validate this candidate. The highest bulk density allowed by the data (based on the upper mass limit and $1\sigma$ lower radius 1.20 \re) is $\rho_{P,max} = 5.10$ \gccc. For highly irradiated super earth candidates such as TOI-125.04 we expect highly irradiated rocky cores with high densities. More observations either with a HARPS-like or more precise instrument such as ESPRESSO \citep{Pepe:2010} would be required to confirm the existence and mass of TOI-125.04. 

For TOI-125.05 (P=13.28 days) we find a radial velocity semi-amplitude consistent with zero; $K=0.2^{+0.4}_{-0.18}$ \ms corresponding to a $2\sigma$ upper mass limit of 2.7 \me. The posterior distribution for the planetary radius presented by \cite{2019AJ....158..177Q} is bi-modal and peaks at 4.2 and 13.5 \re. The $1\sigma$ median for the whole distribution is $8.8^{+4.7}_{-4.4} $\re, which does not reflect the true nature of the posterior. The RV data presented by \citet{2019AJ....158..177Q} and this study both exclude the upper part of the distribution, meaning that if the planet is real it's radius will most likely be similar to that of \Nplanets. We thus only consider the lower part of the radius posterior distribution with 68\% confidence intervals $4.2^{+2.2}_{-1.4}$\re. The highest bulk density allowed by the data (based on the upper mass limit and $1\sigma$ lower radius 2.8 \re) is $\rho_{P,max} = 0.38 $ \gccc. This is a very low density close to being un-physical for a mini-Neptune. We thus conclude that TOI-125.05 is unlikely as a viable planet candidate.

\section{Dynamical stability and system architecture} \label{sec:dyn}

The period ratios in the TOI-125 system are interesting.  If we assume that the low-SNR USP candidate TOI-125.04 is a planet, then the orbital period ratios of adjacent pairs are (beginning from the outside) 2.183, 1.966, and 8.806. The period ratio between planet d and planet c, 2.183, lies at the second most prominent peak in the period ratio distribution \citep{Steffen:2015,Lissauer:2011,Fabrycky:2014} of known exoplanets, close to the 2:1 mean motion resonance (MMR). The origin of this peak is unknown, though it appears both in systems with known intermediate planets (as we see here) and in systems with no observed intermediate planets.

Next, between planets c and b the period ratio is 1.966---sufficiently interior to the 2:1 MMR to be consistent with the observed gap in planet pairs interior to such resonances \citep{Lissauer:2011,Fabrycky:2014}.  There are multiple explanations for this gap interior to the first-order MMR, though none have been demonstrated as the primary cause \citep{petrovich2013,Batygin:2013,Delisle:2012,Chatterjee:2015,Rein:2012,Lithwick:2012}.  Further study of systems like TOI-125 may shed additional light on its origin.  Finally, the innermost planet candidate has an orbital period that is less than one day.  With its neighbour this pair has the largest period ratio in the system. This is consistent with the observed trend that when one member of an adjacent pair of planets has an orbital period less than a day, the period ratio is unusually large \citep{Steffen:2013,Sanchis-Ojeda:2014,Steffen:2016}.  The origin of the ultra-short-period planets remains unknown \citep{2018NewAR..83...37W} though a number of 
hypotheses have been proposed ranging from stripped cores of giant planets \citep{Valsecchi:2014,Konigl:2017} to various dynamical effects coupled with stellar tides \citep{Munoz:2016,Lee:2017,Pu:2019,Petrovich:2019}. The nearby presence of additional small planets would seem not to support the stripped-cores possibility, since hot Jupiter planets tend to be alone with few exceptions \citep{Wright:2009,Steffen:2012,Becker:2015}.  Moreover, \citep{Winn:2017} showed that the metallicity trends of these USP planets do not match those of hot Jupiters --- implying that if USP planets are stripped cores, they must be from smaller, sub-Neptune planets.

The masses of the planets are sufficiently large that in situ formation is unlikely \citep[see eg][]{2014ApJ...795L..15S}.
Thus, formation at larger distances in a protoplanetary disc and migration inwards is a possibility. Planets in resonance are a clear indication of planet migration. 
Furthermore, if the planets formed in the same location in a protoplanetary disc, it would be expected that they would have formed out of similar disc material and thus have similar densities.  The fact that neighbouring planets have significantly different densities is also indicative that they formed in different locations and migrated inwards, as invenstigated for the Kepler~36 system \citep{2012Sci...337..556C} by \cite{2018ApJ...868..138B} and \cite{2018MNRAS.479L..81R}.

\subsection{N-body simulations} \label{sec:nbody}
We attempted to refine the orbital parameters and planet masses for the \Nstar\ system by requiring the system parameters to be compatible with dynamical stability. For this purpose, we considered the 3-planet model for \Nstar\footnote{If real, the USP candidate TOI-125.04 is not expected to play a significant dynamical role in the system, due to its large period ratio with TOI-125b.}, as illustrated in Table \ref{tab:TOI-125}. We used several thousand draws uniformly selected over the full \exofast\ MCMC posterior as 
sets of initial conditions. 

Each set was integrated over a time span of \numprint{5000} years, corresponding to approximately \numprint{91000} revolutions of the outer planet TOI-125d. The simulations were performed with an adaptive timestepping using the N-body 15th-order integrator IAS15 \citep{Rein2015}, available from the software package REBOUND\footnote{The REBOUND code is freely available at \url{http://github.com/hannorein/rebound}.} \citep{Rein2012}. The general relativity correction was included following \citet{Anderson1975}, via the python module REBOUNDx. Then, the stability of each system was explored using the NAFF chaos indicator \citep{Laskar1990,Laskar1993}. The latter consists in estimating precisely the average of the mean motion $n$ of each planet over the first half of the simulation, and repeating this procedure over the second half. The bigger the variation in this average, the more chaotic the system is. Most often, this leads to escapes or close encounters between bodies, defining the system as unstable. Finally, we define a new posterior distribution by keeping only the stable systems. Linking the MCMC exploration of the parameter space with fast chaos indicators is particularly efficient (Stalport et al. in prep). 

\begin{figure}
\includegraphics[width=\columnwidth]{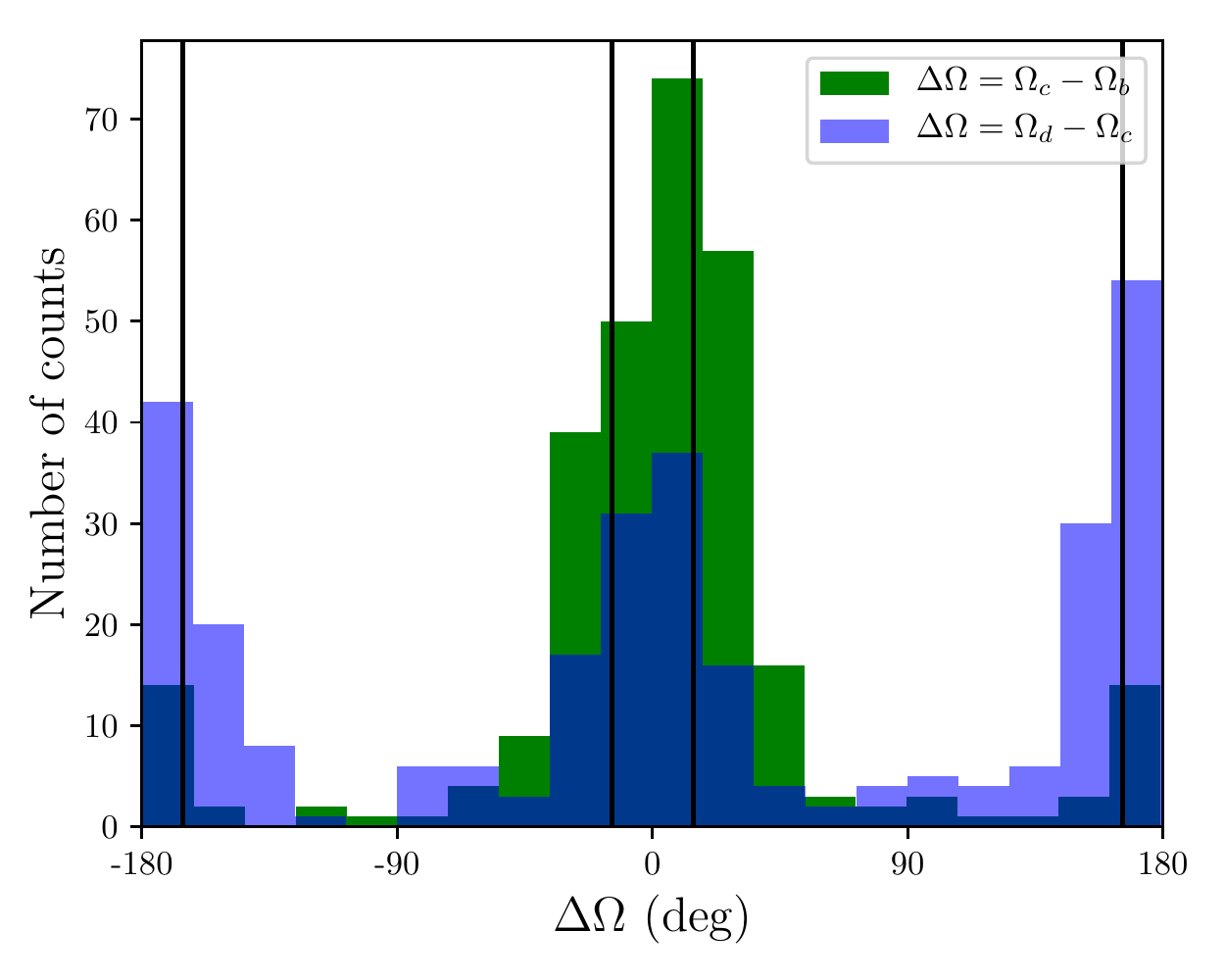} 
\caption{Dynamically stable posterior distribution projected onto $\Omega_c - \Omega_b$ in green, $\Omega_d - \Omega_c$ in blue. The peaks at around $0$ and $\pm 180$ degrees strongly favour the aligned or anti-aligned configurations for the lines of nodes of the planets. \label{DeltaO}}
\end{figure}

The coupled photometric and radial velocity observations give constraints on all the orbital parameters except the longitudes of the nodes of the planets $\Omega$. As a result, this parameter is absent from the \exofast\ MCMC posterior. Therefore, we performed a first series of \numprint{5000} numerical simulations in which the initial values for the $\Omega$ parameters were selected randomly from a uniform distribution between $-\pi$ and $\pi$. The new, dynamically stable posterior distribution strikingly selects only the systems in which the planets have aligned or anti-aligned lines of nodes. This result is illustrated in Figure \ref{DeltaO}. It is explained by the fact that, in these configurations, the mutual inclinations between the adjacent planets are minimal\footnote{The mutual inclination $I_m$ between two orbits is a quantity that depends on the inclination of each orbit $i_k$ and $i_j$ with respect to the plane of the sky, and on the difference in the longitudes of the nodes $\Delta \Omega = \Omega_k - \Omega_j$ ($j$ and $k$ denote the planets). Its expression is $\cos I_{m}~=~\cos i_{k}~\cos i_{j}~+~\cos \Delta\Omega ~\sin i_{k}~\sin i_{j}$.}. Let us note that no information is provided regarding the individual value of $\Omega$ for each planet. However, the dynamical constraints allow us to state that $\Omega_k - \Omega_j = 0$ or $\pi$, for $j$ and $k$ denoting the planets.

Projected onto the other orbital parameters and planetary masses, the dynamically stable posterior distribution does not bring more information. It mimics the original MCMC posterior distribution. This poor refinement can be explained by the aforementioned observation about the lines of nodes. Indeed, many systems turned out to be unstable only because of the unfavourable configurations given by $\Omega$, and the real constraints on the observations were hidden.

\begin{figure}
\includegraphics[width=\columnwidth]{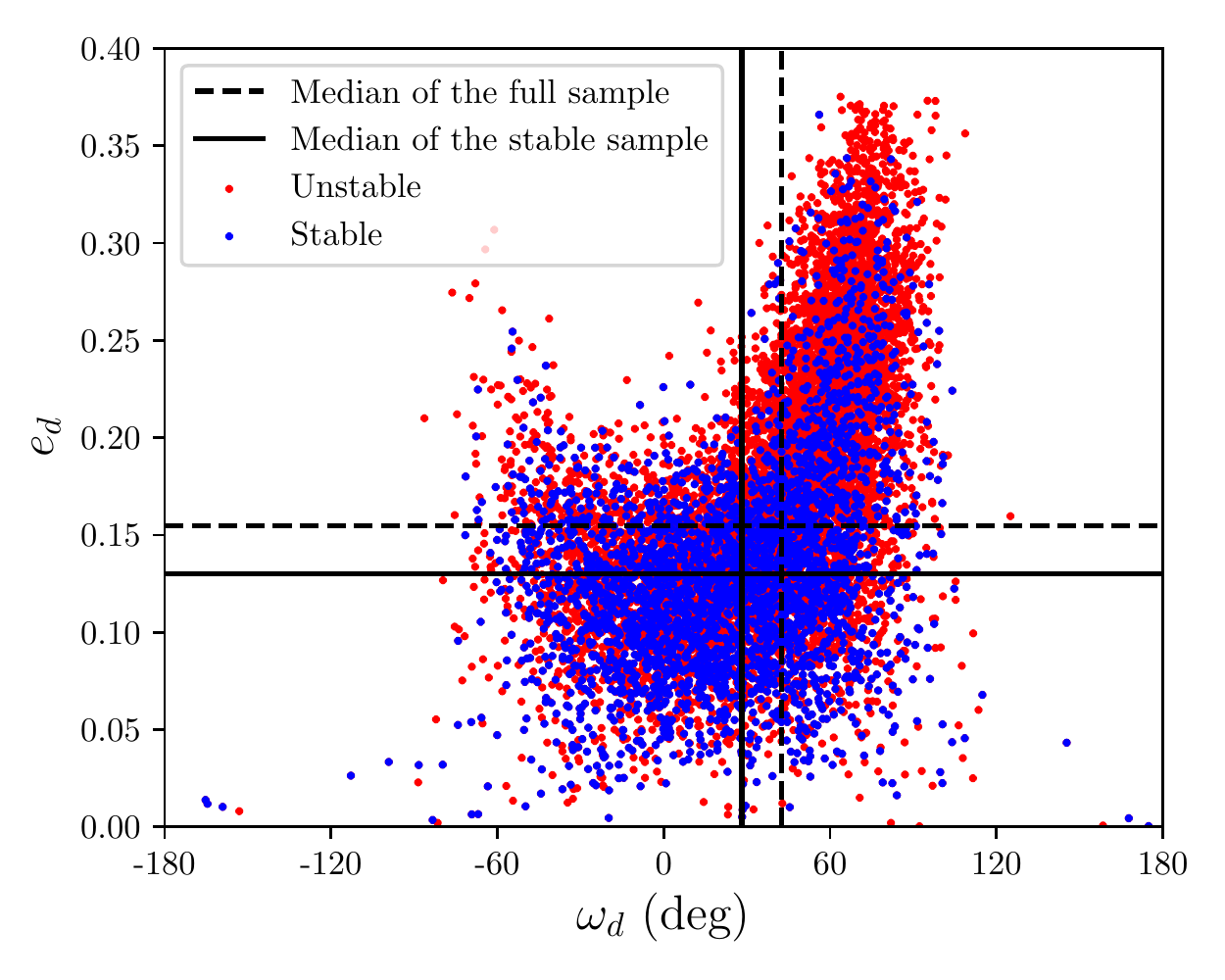} 
\caption{Sample of the posterior distribution from the \exofast\ MCMC of the 3-planets model, projected on the parameters $e_d$ and $\omega_d$. In red, the full sample is projected. The dots are coloured in blue if the corresponding systems are qualified as stable by the NAFF indicator. The black horizontal lines denote the median values of the distributions of $e_d$. The dashed line is associated to the full sample, while the plain line corresponds to the dynamically stable sample. The same applies for $\omega_d$ and the vertical lines. \label{wd-ed}}
\end{figure}

To overcome this bias, we launched a second set of \numprint{10000} numerical simulations. This time, the longitudes of the nodes of the planets were selected randomly in windows around the alignment or anti-alignment, as illustrated by the vertical lines on Figure \ref{DeltaO}. An interesting result of this process is shown in Figure~\ref{wd-ed}. The posterior distribution is projected onto the plane of two parameters, the eccentricity and argument of periastron of the outer planet ($e_d$ and $\omega_d$). As seen in the figure, a branch of solutions at $\omega_d \sim 60^{\circ}$ explores high values of $e_d$. However, this region is disfavoured, as expressed by the decrease in the median of $e_d$.

\begin{figure}
\includegraphics[width=\columnwidth]{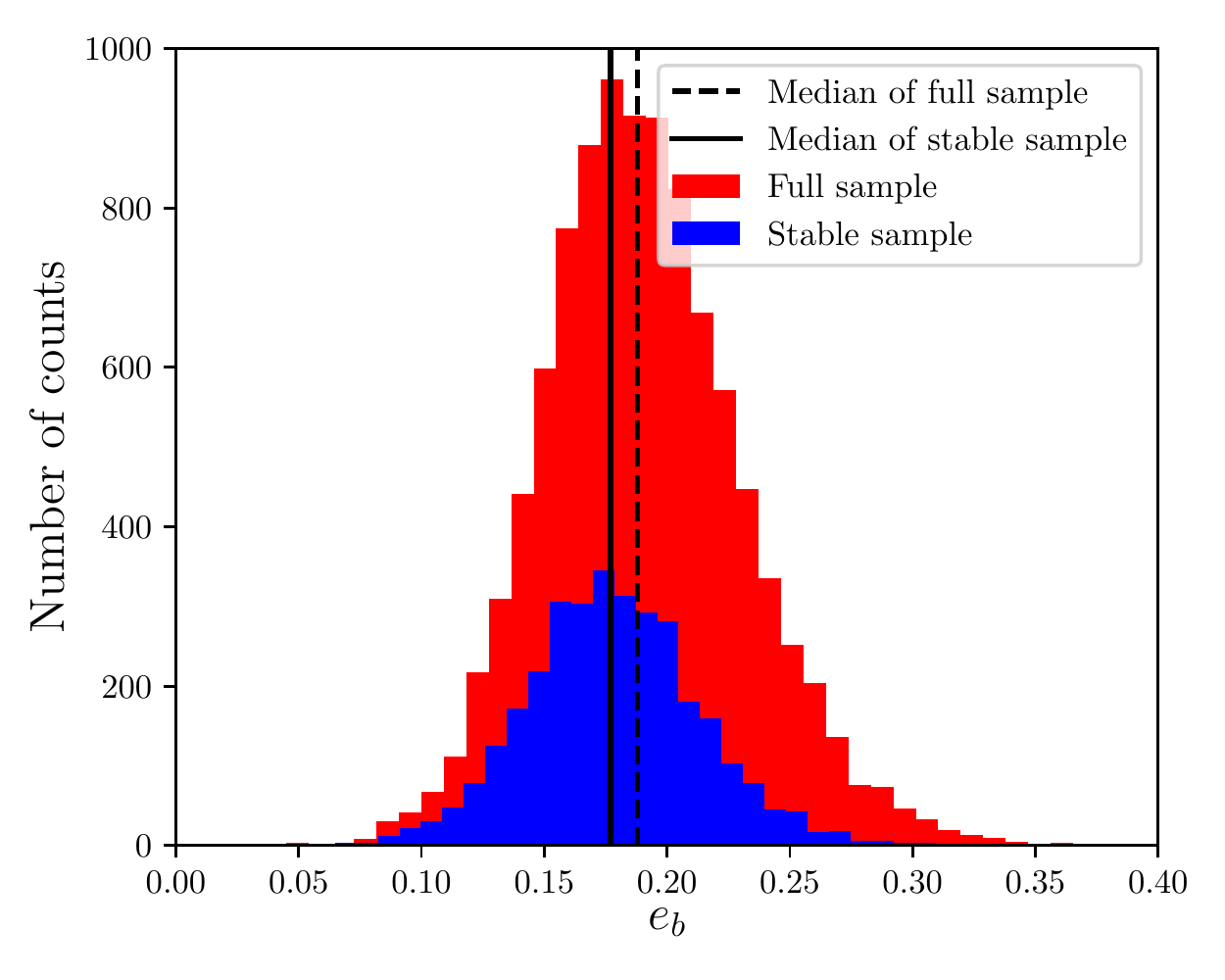} 
\caption{Similar plot as Figure~\ref{wd-ed}. The posterior distribution is now projected onto the single parameter $e_b$. \label{Hist_eb}}
\end{figure}

Another result concerns the relatively high eccentricity of the inner planet, which has a best-fit value of $e_b \sim 0.194$. In Figure~\ref{Hist_eb}, we show the posterior distribution projected onto this parameter in red. The observations are inconsistent with zero eccentricity. A slight displacement towards lower eccentricities is observed in the dynamically stable distribution. Indeed, with the stability constraint, the median of the distribution shifted from $med(e_{b}) \sim 0.188$ (red histogram) to $med(e_{b}) \sim 0.177$ (blue histogram). However, many systems with large eccentricities remain stable. Therefore, such large eccentricities do not seem incompatible with stability.

\subsection{Tidal interactions}
The high eccentricity of planet b also raises questions concerning the tidal evolution of the system.
To investigate those aspects, we also performed N-body integrations taking into account the tidal forces and torques.
To perform those simulations, we used Posidonius\footnote{The Posidonius code is freely available at \url{https://github.com/marblestation/posidonius}.} \citep{2017ewas.confE...8B} which allows to take into account tides, as well as rotational flattening and general relativity using the same prescriptions as in \cite{2015A&A...583A.116B}.

For tides, Posidonius uses an equilibrium tide model \citep{1979M&P....20..301M,1981A&A....99..126H,1998ApJ...499..853E}, for which the tidal dissipation of the different bodies is quantified by the product $k_2 \Delta \tau$ of the constant time lag $\Delta \tau$ and the Love number of degree 2 $k_2$ (the bigger this quantity, the bigger the dissipation and the faster the evolution).
As the underlying assumption of this constant time lag model is that the planet is made of a weakly viscous fluid, it is appropriate for the low-density planets of TOI-125.  
We use a constant time lag similar to Jupiter's \citep[$k_2\Delta \tau \sim 2.5\times 10^{-2}$~s from][]{2010A&A...516A..64L} and explore a range between 1 and $10^2$ times this value.     

Assuming this dissipation for all planets leads to very long evolution timescales. 
In particular, the timescale of circularisation for planet b is about $\gtrsim 10^{10}$~yr and it reaches $10^{13}$~yr for planet d, which is much higher than the estimated age of $\sim 7$~Gyr.
The high eccentricities are therefore not completely surprising and the fact that planet b has not circularised also puts constraints on its dissipation: it cannot be much higher than Jupiter's.
However, the timescales for the damping of the planetary obliquity (angle between the rotation axis and the perpendicular to the orbital plane) and of synchronisation are shorter. 
Assuming the same dissipation as Jupiter, and even assuming the lower estimate of the age (2.5 Gyr), we find that planets b and c should have a damped obliquity (less than a few degrees) and an evolved rotation.
In our model, the evolved rotation period is the pseudo-synchronisation period, which depends on eccentricity \citep{1981A&A....99..126H}. 
Depending on the age of the system, the obliquity and rotation of planet d might still be evolving: if the system is older than $\sim$6~Gyr, the obliquity should be very small and the rotation should be very close to the pseudo-synchronisation rotation.

Of course, there is a strong uncertainty on the dissipation factor of planets, these planets could dissipate more energy than what is estimated for Jupiter (with processes such as tidal inertial waves in the convective region \citealt{2004ApJ...610..477O}).
But unless the age of the system is close to its upper estimate of 11~Gyr, the fact that planet b still has a high eccentricity tends to indicate that dynamical tide processes are not very efficient.




\section{Internal structure} \label{sec:internal}

In order to characterise the internal structure of \Nplanets\ we construct models considering a pure iron core, a silicate mantle, a pure water layer and a H-He atmosphere. The models follow the basic structure model of \cite{2017A&A...597A..38D}, with the equation of state (EOS) for the iron core taken from \cite{2018Icar..313...61H}, and the EOS of the silicate-mantle from \cite{2009GGG....1010014C}. For water we use the quotidian EOS of \cite{2013MNRAS.434.3283V} for low pressures and the one of \cite{2007ApJ...669.1279S} for pressures above 44.3 GPa. The hydrogen-helium (H-He) EOS is SCVH \citep{1995ApJS...99..713S} assuming a proto-solar composition. We then use a generalised Bayesian inference analysis using a Nested Sampling scheme \citep[e.g.][]{Buchner2016}. 
We then quantify the degeneracy between interior parameters and produce posterior probability distributions. The interior parameters that are inferred include the masses of the pure-iron core, silicate mantle, water layer and H-He atmospheres. For this analysis we use the stellar Fe/Si and Mg/Si ratios from Table \ref{tab:spec} as a proxy for the planet abundandances.

\begin{figure}
    \includegraphics[width=\columnwidth]{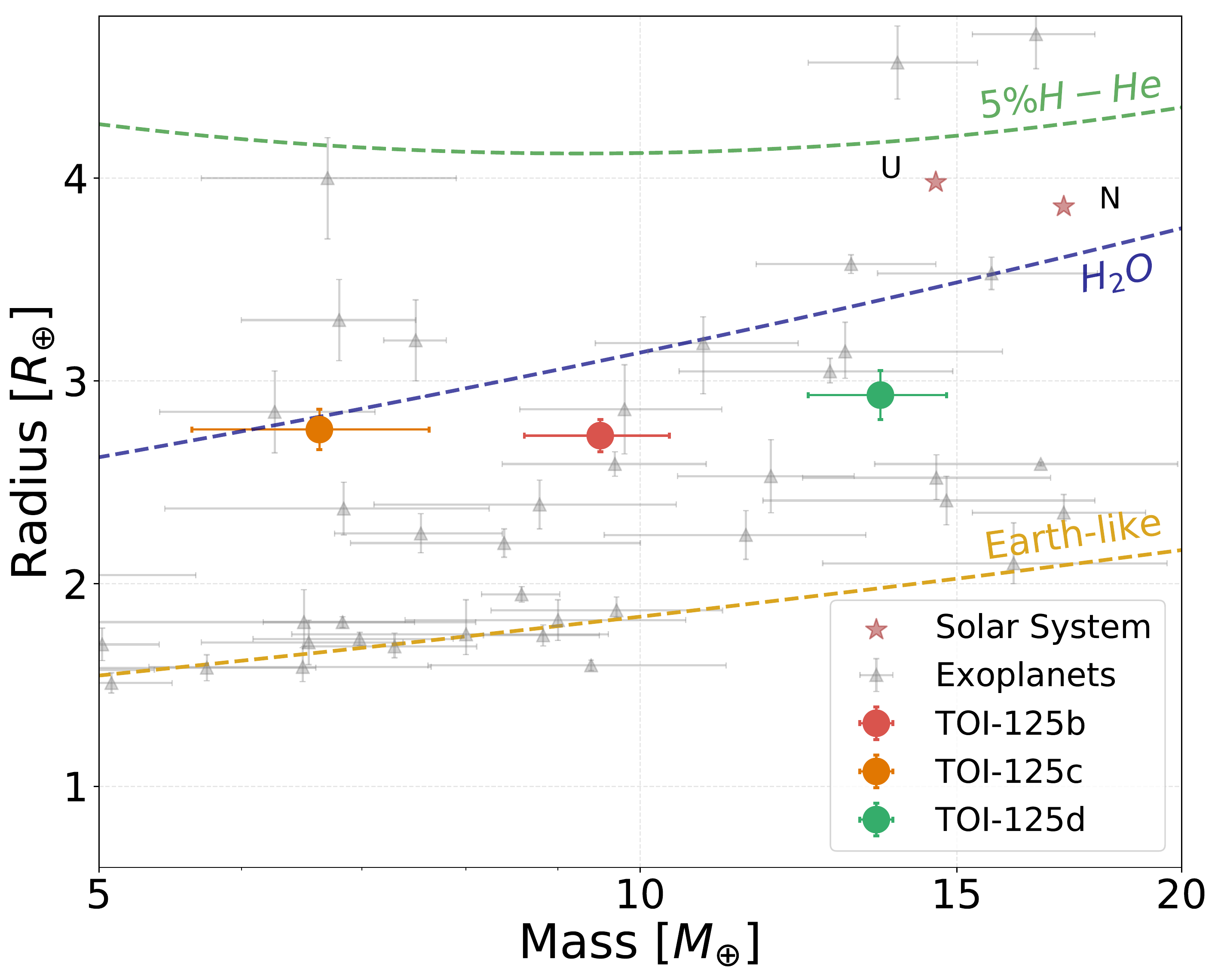}
  \caption{Mass-radius diagram of exoplanets with accurate mass and radius determination \citep{Otegi:2019}. Also shown are the composition lines of an Earth-like planet, pure-water and 95\% $\mathrm{H_2 O+5\% H-He}$. \label{fig:MR}}  
\end{figure}

Figure~\ref{fig:MR} shows the mass-radius relation for a pure-water curve and a planet with 95\% water and 5\% H-He atmosphere subjected to a stellar radiation of $F/F_{\oplus}=100$ (comparable to the case of the TOI-125 planets). 
All three planets could in principle either consist of a rocky core with a massive water envelope (mostly in the form of supercritical steam) or a rocky core with a likely high metallicity H-He envelope (up to 5\% in mass of H-He). The position of the three planets in the insolation radius diagram (Figure \ref{fig:insolation}), above the evaporation valley \citep{Fulton:2017,Fulton:2018,2018MNRAS.479.4786V} indicate however that the latter scenario (i.e. involving a H2/He envelope) is the most plausible one \citep{2017ApJ...847...29O, 2018MNRAS.476..759G}. Spectroscopic transit measurements will hopefully help to discriminate between the two aforementioned cases owing to the relative proximity of the TOI-125 system, see Section \ref{sec:atmos} for a more in-depth discussion. Transit observations of the exoplanet GJ1214b -- which lies in a somewhat similar insolation radius mass parameter space than TOI-125 planets -- have however shown that clouds may limit our ability to conclude on the true nature of these objects \citep{2014Natur.505...69K}.


Table \ref{tab:internal} lists the inferred mass fractions of the core, mantle, water layer and H-He atmosphere from our structure models. We find median H-He mass fractions of 2.3\% for TOI-125b, 2.9\% for TOI-125c, and 4.5\% for TOI-125d. These estimates are lower bounds since structure models considering H-He envelopes enriched with heavy elements could result in even higher values. This is because enriched H-He atmospheres  are more compressed, and can therefore increase the planetary H-He mass fraction.  Indeed, formation models of mini-Neptunes suggest that forming such planets without envelope enrichment is very unlikely \citep{2017ApJ...848...95V}. 

TOI-125b and TOI-125c are expected to have very similar compositions, with core and water layer mass fractions of $\sim 30\%$ and a mantle mass fraction of $\sim 40\%$. TOI-125d, instead, has a slightly higher water mass fraction of 35\%, and a smaller fraction of refractory materials with a core mass fraction of 26\% and mantle mass fraction of 35\%. 


\begin{table}
\caption{Inferred interior structure properties of \Nplanets. \label{tab:internal}}
\begin{tabular}{lccc}
\hline
 Interior Structure& TOI-125b& TOI-125c& TOI-125d\\[1pt]
\hline 
 $M_{\mathrm{core}}/M_{\mathrm{total}}$& $0.31 ^{+0.18} _{-0.32}$&$0.31 ^{+0.16} _{-0.27}$&$0.26 ^{+0.16} _{-0.21}$\\[4pt]
   $M_{\mathrm{mantle}}/M_{\mathrm{total}}$& $0.39
   ^{+0.17} _{-0.26}$ & $0.38 ^{+0.18} _{-0.29}$ & $0.36 ^{+0.18} _{-0.31}$\\[4pt]
  $M_{\mathrm{water}}/M_{\mathrm{total}}$& $0.32 ^{+0.20} _{-0.24}$ & $0.32 ^{+0.17} _{-0.24}$ & $0.36 ^{+0.16} _{-0.21}$\\[4pt]
  $M_{\mathrm{H-He}}/M_{\mathrm{total}}$&  $0.020
  ^{+0.006} _{-0.008}$ & $0.027 ^{+0.007} _{-0.010}$ & $0.041 ^{+0.009} _{-0.012}$\\[4pt]
  \hline

\end{tabular}
\label{tab:internal_structure}
\end{table}

\section{Potential for atmospheric characterisation} \label{sec:atmos}

 \begin{figure}
 \label{fig:insolation} 
    \includegraphics[width=\columnwidth]{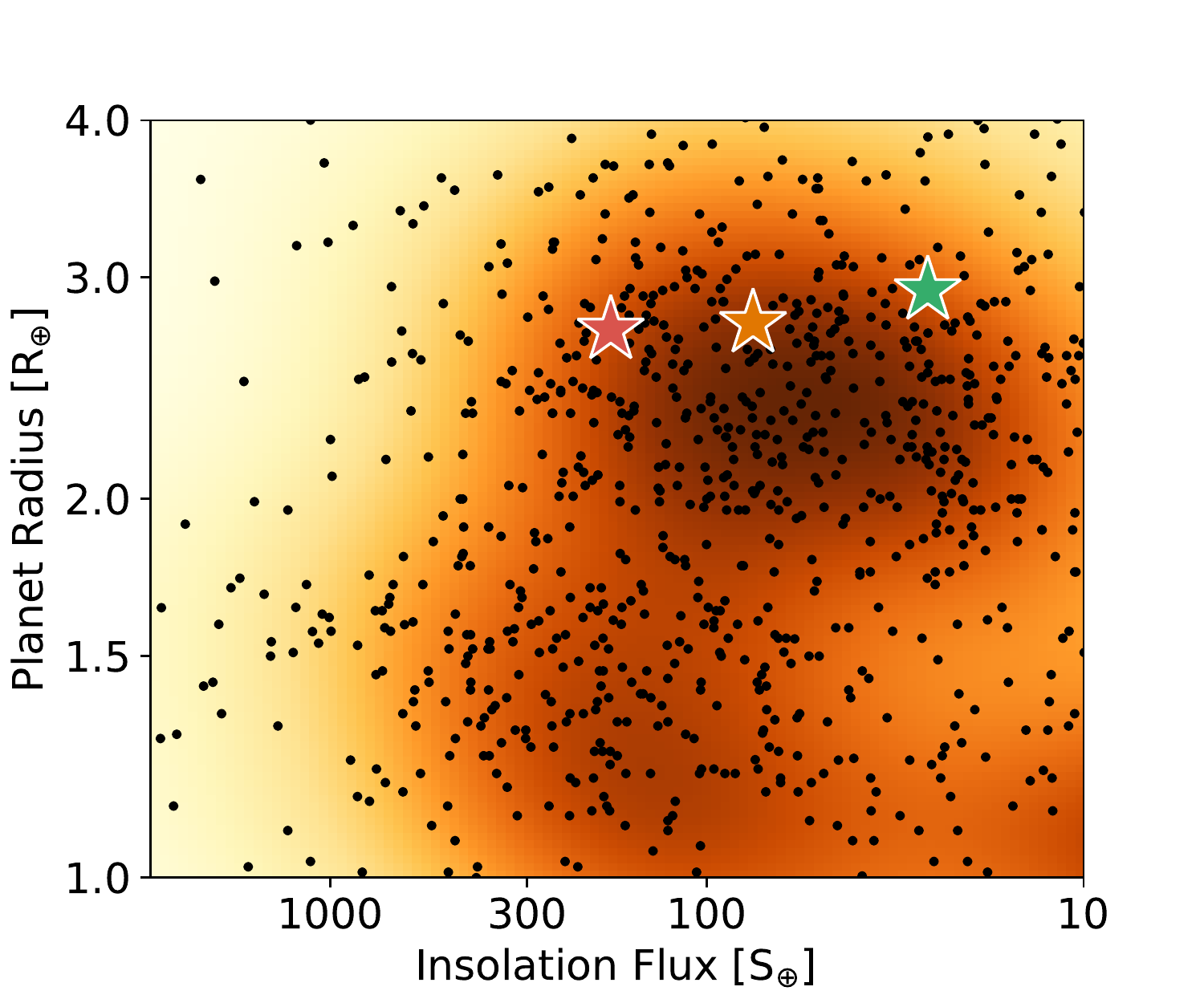}
  \caption{Insolation flux relative to Earth plotted against radii for known exoplanets extracted from NASA Exoplanet Archive, as presented in \citet{Fulton:2017,Fulton:2018}. The orange contours indicate point density (not occurrence), showing the separate populations of mini-Neptunes and super-Earths. \Nplanets\ are plotted as three stars in the same colours as in Figure~\ref{fig:LCs}, \ref{fig:RVs} and \ref{fig:MR}. }

\end{figure}

Our analysis of the internal structure (see Table \ref{tab:internal_structure}), as well as the position of the three planets in the insolation-radius diagram (see Figure~\ref{fig:insolation}), indicate that all three planets might have a water dominated atmosphere with a small contribution from lighter elements at the order of a few percent. If these light elements are evaporated over time (especially for TOI-125b, the most irradiated in the system), their observation could be used to study the planets' exospheres.

Due to the significant distance of the system ($111.40$~pc), the absorption of the interstellar medium (ISM) puts Lyman-$\alpha$ observations out of reach. However, H-alpha and HeI, which do not suffer from ISM absorption, can be used to detect a potential escaping planetary outflow. H-alpha, and other Balmer series lines, have been detected for several exoplanets, showing deep absorption features observed at high spectral resolution \citep{Jensen:2012, Cauley:2017, Jensen:2018, Yan:2018}. Likewise, the well known HeI triplet in the infrared \citep{Seager:2000, Oklopcic:2018, Oklopcic:2019}, has also successfully detected exospheric absorption in other systems \citep{Allart:2018, Nortmann:2018, Salz:2018, Allart:2019}.

The possible water rich composition from Table \ref{tab:internal_structure} could be verified via observations in the infrared, and thus provide valuable insights into the water composition in a multi planet system with three similarly sized planets but different masses and insolations. However, observations from the ground are challenging due to the planets' sizes and observational windows. We estimated that one transit observations would not be useful to detect water bands for TOI-125b (scale height $38$ km) with NIRPS at the ESO 3.6m telescope \citep{Bouchy:2017}. Observing multiple transits would require a dedicated large program spanning several years given the possible observational windows from Chile. It is, however, a prime target for observations with the next generation of ELTs, particularly with the HIRES optical-to-NIR spectrograph at the E-ELT \citep{Marconi:2016} and CRIRES+ at the VLT \citep{Follert:2014}. 

Using the Pandexo Exposure Time Calculator for \emph{HST}\footnote{\footnotesize{Available at \url{https://exoctk.stsci.edu/pandexo/}.}}, we estimate that the precision with which we can measure the transmission spectrum of TOI-125b using the Wide-Field Camera 3 (WFC3) instrument, in five transits, is $\sim$30 ppm near the 1.4 $\mu$m water feature. The expected water signature at 5-scale heights has a depth of approximately 20 ppm, thus detecting this feature with \emph{HST} would be challenging for a planet with an atmosphere as compact as TOI-125b. However, all three planets are prime targets for JWST's NIRSpec.


\section{Conclusions}
We confirm the detection of three mini-Neptunes around \Nstar\ found by \tess\ using HARPS RV measurements. \Nplanets\ have all have similar radii; $2.726 \pm 0.075$~\re, $2.759 \pm 0.10$~\re\ and $2.93 \pm 0.17$~\re, respectively. The three planets differ greatly in mass however with $ 9.50 \pm 0.88$~\me, $ 6.63 \pm 0.99$~\me\ and $13.6 \pm 1.2$~\me, yielding a high-low-higher pattern in terms of density when moving outward in the system. For the two marginal planet candidates TOI-125.04 and TOI-125.05 we derive 2-$\sigma$ upper mass limits of 1.6~\me\ and 2.7~\me, respectively. For TOI-125.05 this mean it is unlikely as a viable planet candidate.

The system exhibit an intriguing architecture with the two inner planet slightly interior to the 2:1 MMR while the two outer planets are slightly external to the 2:1 MMR. TOI-125b and TOI-125d both show significant orbital eccentricities. We analyse the dynamics of the system using N-body simulations and demonstrate that planetary orbits are stable despite the high eccentricities. Based on N-body simulations coupled with tidal forces and torques we conclude that the dynamical tide processes cannot be very efficient in order for TOI-125b to retain it's high eccentricity of $e_b = 0.194 ^{+0.041}_{-0.036}$.

Our analysis of the internal compositions of these three planets yield that they all most likely retain  H-He atmospheres and a significant water layer which could be detected though transmission spectroscopy. This is expected for planets sitting on top of the radius gap (see Figure \ref{fig:insolation}), receiving less than 300 times the stellar insolation than that of the Earth.


\section*{Acknowledgements}
We thank the anonymous referee for providing thoughtful comments that allowed us to improve on this paper.
This study is based on observations collected at the European Southern Observatory under ESO programmes 0101.C-0829, 1102.C-0249, 1102.C-0923, 0102.C-0525 and 0102.C-0451.\\ 
We  thank  the  Swiss  National  Science  Foundation  (SNSF) and the Geneva University for their continuous support to our planet search programs. This work has been in particular carried out in the frame of the National Centre for Competence in Research {\it PlanetS} supported by the Swiss National Science Foundation (SNSF). \\ 
This publication makes use of The Data \& Analysis Center for Exoplanets (DACE), which is a facility based at the University of Geneva (CH) dedicated to extrasolar planets data visualisation, exchange and analysis. DACE is a platform of the Swiss National Centre of Competence in Research (NCCR) PlanetS, federating the Swiss expertise in Exoplanet research. The DACE platform is available at \url{https://dace.unige.ch}. \\ 
This paper includes data collected by the \tess\ mission. Funding for the \tess\ mission is provided by the NASA Explorer Program. 
Resources supporting this work were provided by the NASA High-End Computing (HEC) Program through the NASA Advanced Supercomputing (NAS) Division at Ames Research Center for the production of the SPOC data products.\\ 
This work has made use of data from the European Space Agency (ESA) mission \gaia\ (\url{https://www.cosmos.esa.int/gaia}), processed by the \gaia\ Data Processing and Analysis Consortium (DPAC, \url{https://www.cosmos.esa.int/web/gaia/dpac/consortium}). Funding for the DPAC has been provided by national institutions, in particular the institutions participating in the \gaia\ Multilateral Agreement. \\ 
This research has made use of the NASA Exoplanet Archive, which is operated by the California Institute of Technology, under contract with the National Aeronautics and Space Administration under the Exoplanet Exploration Program. \\
DJA acknowledges support from the STFC via an Ernest Rutherford Fellowship (ST/R00384X/1). 
The IA/Portuguese team was supported by FCT/MCTES through national funds and by FEDER - Fundo Europeu de Desenvolvimento Regional through COMPETE2020 - Programa Operacional Competitividade e Internacionaliza\c{c}\~{a}o by these grants: UID/FIS/04434/2019; PTDC/FIS-AST/32113/2017 \& POCI-01-0145-FEDER-032113; PTDC/FIS-AST/28953/2017 \& POCI-01-0145-FEDER-028953. 
VA acknowledges the support from FCT through Investigador FCT contract nr. IF/00650/2015/CP1273/CT0001. 
 SH acknowledge support by the fellowships PD/BD/128119/2016 funded by FCT (Portugal). 
SCCB acknowledges support from FCT through Investigador FCT contracts IF/01312/2014/CP1215/CT0004.
O.D.S.D. acknowledges the support from FCT (Portugal) through work contract DL 57/2016/CP1364/CT0004. 
MRD acknowledges support of CONICYT-PFCHA/Doctorado Nacional-21140646 and Proyecto Basal AFB-170002. 
JSJ acknowledges support from FONDECYT grant 1161218. 
FM acknowledges support from The Royal Society Dorothy Hodgkin Fellowship. 
JVS and LAdS are supported by funding from the European Research Council (ERC) under the European Union's Horizon 2020 research and innovation programme (project {\sc Four Aces}; grant agreement No 724427).  
DJAB acknowledges support from the UK Space Agency. 
JNW acknowledges support from the Heising-Simons Foundation. 
NN is supported by JSPS KAKENHI Grant Numbers JP18H01265 and JP18H05439, and JST PRESTO Grant Number JPMJPR1775.
CD acknowledges support from the Swiss National Science Foundation under grant PZ00P2\_174028.
KWFL acknowledge support by DFG grants RA714/14-1 within the DFG Schwerpunkt SPP 1992, "Exploring the Diversity of Extrasolar Planets". 
DB and JLB habe been funded by the Spanish State Research Agency (AEI) Projects No.ESP2017-87676-C5-1-R and No. MDM-2017-0737 Unidad de Excelencia Mar{\'{\i}}a de Maeztu Centro de Astrobiolog{\'{\i}}a (CSIC-INTA). 
SM acknowledges support from the Spanish Ministry under the Ramon y Cajal fellowship number RYC-2015-17697.  
R.A.G. acknowledges the support from PLATO and GOLF CNES grants. 
This project has received funding from the European Union's Horizon 2020 research and innovation program under the Marie Sklodowska-Curie Grant Agreement No. 832738/ESCAPE. M.T. acknowledges funding from the Gruber Foundation.
M.F., I.G. and C.M.P. gratefully acknowledge the support of the Swedish National Space Agency (DNR 163/16 and 174/18). 



\bibliographystyle{mnras}
\bibliography{TOI-125_HARPS} 



\appendix

\onecolumn
\section{Author affiliations} \label{sec:affil}
$^{1}$Geneva Observatory, University of Geneva, Chemin des Mailettes 51, 1290 Versoix, Switzerland\\
$^{2}$Dipartimento di Fisica, Universita degli Studi di Torino, via Pietro Giuria 1, I-10125, Torino, Italy\\
$^{3}$Centre for Exoplanets and Habitability, University of Warwick, Gibbet Hill Road, Coventry, CV4 7AL, UK\\
$^{4}$Department of Physics, University of Warwick, Gibbet Hill Road, Coventry CV4 7AL, UK \\
$^{5}$Departamento de Astronom\'ia, Universidad de Chile, Camino El Observatorio 1515, Las Condes, Santiago, Chile\\
$^{6}$Leiden Observatory, P.O. Box 9513, NL-2300 RA Leiden, The Netherlands\\
$^{7}$Department of Space, Earth and Environment, Chalmers University of Technology, Onsala Space Observatory, SE-439 92, Sweden\\
$^{8}$Instituto de Astrof\'isica e Ci\^encias do Espa\c{c}o, Universidade do Porto, CAUP, Rua das Estrelas, 4150-762 Porto, Portugal\\
$^{9}$Departamento de F\'isica e Astronomia, Faculdade de Ci\^{e}ncias, Universidade do Porto, Rua do Campo Alegre, 4169-007 Porto, Portugal\\
$^{10}$Division of Geological and Planetary Sciences, California Institute of Technology, 1200 East California Blvd, Pasadena, CA, USA 91125\\
$^{11}$Instituto de Astrofisica de Canarias, Tenerife, Spain\\ 
$^{12}$Dpto. Astrof\'{i}sica, Universidad de La Laguna, Tenerife, Spain \\
$^{13}$Department of Physics and Astronomy, University of Nevada, Las Vegas, Las Vegas, NV, 89154\\
$^{14}$Th{\"u}ringer  Landessternwarte  Tautenburg,  Sternwarte  5,  D-07778 Tautenburg, Germany\\
$^{15}$ Max-Planck-Institut f\"ur Astronomie, K\"onigstuhl 17, 69117 Heidelberg, Germany\\
$^{16}$Institute for Computational Science, University of Zurich, Winterthurerstr. 190, CH-8057 Zurich, Switzerland\\
$^{17}$European Southern Observatory (ESO), Alonso de C\'ordova 3107, Vitacura, Casilla 19001, Santiago de Chile\\
$^{18}$Department of Physics and Kavli Institute for Astrophysics and Space Research, MIT, Cambridge, MA 02139, USA\\
$^{19}$Center for Astrophysics | Harvard \& Smithsonian, 60 Garden Street, Cambridge, MA 02138, USA\\
$^{20}$Department of Earth, Atmospheric, and Planetary Sciences, MIT, 77 Massachusetts Avenue, Cambridge, USA\\
$^{21}$Department of Aeronautics and Astronautics, MIT, 77 Massachusetts Avenue, Cambridge, MA 02139, USA\\
$^{22}$Department of Astrophysical Sciences, Princeton University, 4 Ivy Lane, Princeton, NJ 08544, USA\\
$^{23}$NASA Ames Research Center, Moffett Field, CA 94035, USA\\
$^{24}$Depto. de Astrof\'isica, Centro de Astrobiolog\'ia (CSIC-INTA), ESAC campus 28692 Villanueva de la Ca\~nada (Madrid), Spain\label{cab}\\
$^{25}$Aix Marseille Univ, CNRS, CNES, LAM, Marseille, France\\
$^{26}$Astrophysics Science Division, NASA Goddard Space Flight Center, Greenbelt, MD 20771\\
$^{27}$McDonald Observatory and Department of Astronomy, University of Texas, Austin TX, USA\\
$^{28}$Universidad de Buenos Aires, Facultad de Ciencias Exactas y Naturales. Buenos Aires, Argentina.\\
$^{29}$CONICET - Universidad de Buenos Aires. Instituto de Astronom\'{i}a y F\'isica del Espacio (IAFE). Buenos Aires, Argentina.\\
$^{30}$IRFU, CEA, Universit\'e Paris-Saclay, Gif-sur-Yvette, France\\
$^{31}$AIM, CEA, CNRS, Universit\'e Paris-Saclay, Universit\'e Paris Diderot, Sorbonne Paris Cit\'e, F-91191 Gif-sur-Yvette, France\\
$^{32}$Rheinisches Institut f\"ur Umweltforschung an der Universit\"at zu K\"oln, Aachener Strasse 209, 50931 K\"oln Germany\\
$^{33}$Centre for Astronomy and Astrophysics, Technical University Berlin, Hardenbergstrasse 36, 10585 Berlin, Germany\\
$^{34}$Department of Astronomy, University of Tokyo, 7-3-1 Hongo, Bunkyo-ky, Tokyo 113-0033, Japan\\
$^{35}$Astrobiology Center, 2-21-1 Osawa, Mitaka, Tokyo 181-8588, Japan\\
$^{36}$JST, PRESTO, 2-21-1 Osawa, Mitaka, Tokyo 181-8588, Japan\\
$^{37}$National Astronomical Observatory of Japan, 2-21-1 Osawa, Mitaka, Tokyo 181-8588, Japan\\
$^{38}$Center for Space and Habitability, University of Bern, Gesellschaftsstrasse 6, 3012 Bern, Switzerland\\
$^{A}$Institute of Planetary Research, German Aerospace Center (DLR), Rutherfordstrasse 2, D-12489 Berlin, Germany\\
$^{C}$Institute of Geological Sciences, FU Berlin, Malteserstr. 74-100, D-12249 Berlin\\
$^{41}$Astronomy Department and Van Vleck Observatory, Wesleyan University, Middletown, CT 06459, USA\\
$^{42}$NASA Sagan Fellow, Department of Astronomy, Unversity of Texas at Austin, Austin, TX, USA\\
$^{43}$Mullard Space Science Laboratory, University College London, Holmbury St Mary, Dorking, Surrey RH5 6NT, UK


\bsp	
\label{lastpage}
\end{document}